\documentstyle[epsf,10pt]{article}

\renewcommand{\o}{\over}
\newcommand{\p}{\partial}
\newcommand{\be}{\begin{equation}}
\newcommand{\ee}{\end{equation}}
\newcommand{\bea}{\begin{eqnarray}}
\newcommand{\eea}{\end{eqnarray}}

\newcommand{\bp}{{\bf B}_\rp}
\newcommand{\vp}{{\bf v}_\rp}
\newcommand{\efi}{{\bf e}_\phi}
\newcommand{\ra}{\varpi_{\rm A}}
\newcommand{\po}{\varpi}
\newcommand{\rd}{{\rm d}}
\newcommand{\rp}{{\rm p}}
\newcommand{\rs}{{\rm s}}
\newcommand{\pr}{\prime}
\newcommand{\gapprox}{\;\rlap{\lower 2.5pt
             \hbox{$\sim$}}\raise 1.5pt\hbox{$>$}\;}
\newcommand{\lapprox}{\;\rlap{\lower 2.5pt
             \hbox{$\sim$}}\raise 1.5pt\hbox{$<$}\;}
\newcommand{\bfg}[1]{\setbox0=\hbox{#1}%
  \kern-.025em\copy0\kern-\wd0
  \kern.05em\copy0\kern-\wd0
  \kern-.025em\raise.0433em\box0}

\newcommand{\aaa}[1]{{\em Astron.\ Astrophys.,} {\bf #1}}

\newcommand{\annrev}[1]{{\em Ann.\ Rev.\ Astron.\ Astrophys.,} {\bf #1}}
\newcommand{\apj}[1]{{\em Astrophys. J.,} {\bf #1}}
\newcommand{\aj}[1]{{\em Astron. J.,} {\bf #1}}

\newcommand{\apjs}[1]{{\em Astrophys.\ J.\ Suppl.,} {\bf #1}}

\newcommand{\mnras}[1]{{\em Mon.\ Not.\ R.\ astron.\ Soc.,} {\bf #1}}
\newcommand{\nature}[1]{{\em Nature,} {\bf #1}}

\newcommand{\pasj}[1]{{\em Publ.\ Astr.\ Soc.\ Japan,} {\bf #1}}

\renewcommand{\sp}[1]{{\em Solar Phys.,} {\bf #1}}

\renewcommand{\index}[1]{}
\newcommand{\tab}[1]{table~\ref{#1}}
\newcommand{\fig}[1]{figure~\ref{#1}}
\newcommand{\sect}[1]{section~\ref{#1}}
\newcommand{\eq}[1]{equation~\ref{#1}}

\begin{document}
\centerline{\Large Magnetohydrodynamic Jets and Winds}
\centerline{\Large from Accretion Disks}
\vspace{0.5 cm}
\centerline{H.C. SPRUIT}
\centerline{\em Max-Planck-Institut f\"ur Astrophysik}
\centerline{Postfach 1523, D-85740 Garching, Germany}
\vspace{0.5 cm}

{\small
The theory of magnetically accelerated outflows and jets from accretion disks
is reviewed at an introductory level, with special attention to problem areas
like the launching conditions of the flow at the disk surface, stability of the
magnetic field, and collimation mechanisms. This text will appear in R.A.M.J. Wijers,
M.B. Davies and C.A. Tout, eds., {\it Physical processes in Binary Stars}, Kluwer
Dordrecht, 1996 (NATO ASI series).
}
\vspace{1 cm}

\section{Introduction: the case for magnetic acceleration}

\index{jets}\index{outflows}
Narrow, high speed outflows (jets) and less well collimated `bipolar' outflows
are observed from very different cosmical objects, ranging from protostars in
the solar neighborhood, to galactic X-ray binaries, to the nuclei of
active galaxies. The magnetic acceleration mechanism for outflows from
accretion disks has gained significant popularity as an explanation for each
of these forms of outflow. The model can account for high speeds (for example,
Lorentz factors of 10 or higher in AGN and galactic black hole binaries), high
degrees of collimation, and large momentum fluxes. Though other processes can
also, to varying degrees, account for these properties (e.g.\ Blandford 1993),
the magnetic model combines them in a natural way. While some processes are
very different in protostellar disks and the AGN or X-ray binary disks, in
particular those that produce the observed radiation, the physics of the
magnetic acceleration model is to a large degree independent of these.
Progress made by development of the model for explaining observations in one
area is therefore likely to have impact for the interpretation of other
outflows as well.

In spite of the rather general applicability of the magnetic mechanism, it is
unlikely that it is involved in all cases. There are a number of bipolar
looking objects in the sky where the explanation may well be purely
hydrodynamical, such as the outflows in $\eta$ Carinae, and in particular the
planetary nebulae (Icke et al. 1992).

The protostellar outflows play a key role in supporting the
magnet\-ic/centri\-fugal model. The momentum flux in these objects can be
measured from flow speeds and inferred mass densities, and in many cases turns
out to be much larger than can be accounted for by the nearest competing
mechanism, radiation
pressure from the central star (for a review see K\"onigl and Ruden, 1993).
The protostellar outflows are also the ones in which observations are most
likely, in the near future, to reveal their inner workings by direct imaging,
as shown in \tab{hsta1}. See the contributions elsewhere in this volume
for more about this subject.

\begin{table}[bp]
\caption
{\label{hsta1}Angular size of the accelerating region (assumed to
be 100 times the typical size $r_0$ of the inner disk) for
jet-producing objects.}
\begin{tabular}{lccc}
 & inner disk & distance & angular scale ($"$) \\
 &  $r_0$ & D & $100\,r_0/D$ \\
 nearby protostar & $3 R_\odot$ & 500 pc & 0.003 \\
 nearby active galactic nucleus& 100 AU & 10 Mpc & 0.001 \\
 galactic black hole candidate& 100 km & 2 kpc & $3\,10^{-8}$
\end{tabular}
\end{table}

\section{Presence of jets and outflows in binaries}
\label{hsobs}\index{X-ray Binaries!jets}
Jets are now known from all classes of X-ray binaries, i.e. mass transfering
binaries (Lewin et al. 1995) in which the primary is a black hole or neutron
star. (Ignoring here the `supersoft' X-ray binaries in which the primaries
appear to be white dwarfs). Among the massive X-ray binaries (HMXB) in which
the companion is an early type star, there are the well known jets of SS 433,
and the radio jets from Cyg X-3 (Strom, van Paradijs and van der Klis, 1989)
and the galactic center source 1E140.7 - 2942 (Mirabel et al. 1992). Among the
low mass X-ray binaries (LMXB) with neutron star primaries, a jet is known
only from Cir X-1 (Stewart et al. 1993). Until recently no jets were known
from LMXB with black hole candidate primaries, but this has changed with the
discovery of superluminal jets in the variable source GRS 1915+105 (Mirabel
1994), and the transient GRO 1655-40 (= X-ray Nova Sco 1994), (Hjellming and
Rupen 1995). The physics in the inner disks of these objects must be rather
similar to that in the central engines of AGN (e.g.\ Begelman, Blandford and
Rees 1984), and it is pleasing to see that they can produce very similar jets,
though on a much smaller scale (for more observational similarities between
galactic black hole candidates and AGN see Sunyaev et al. 1991 and references
therein).

Outflows without evidence for jets exist in Cataclysmic
Variables\index{Cataclysmic Variables!outflows} (CV:
binaries transfering mass from a main sequence star to a white dwarf, see Hack
and La Dous, 1993). P-Cygni profiles indicating mass loss are seen in UV lines
in Dwarf Novae (DN) when in outburst (e.g.\ Drew and Verbunt, 1988). No
evidence for outflow is known for DN in quiescence. In UX Uma systems (CV
with steady mass transfer), evidence for mass loss comes from single-peaked
line profiles (in contrast with the classical double-peaked profiles of
accretion disks), and the presence of additional `uneclipsed light' in
eclipsing systems (`SW Sex syndrome', Thorstensen et al. 1991). It is fair to
say that no collimated outflow has yet been observed from a CV. One might
wonder if this has something to do with the fact that the primaries are white
dwarfs, but the case of R Aqr\index{R Aqr} shows that this cannot be the
case. R Aqr is a
white dwarf with a Mira type giant companion in a 44 yr orbit; mass transfer
occurs because of the dense stellar wind from the Mira. It has a jet which is
visible at optical as well as radio wavelengths (Burgarella and Paresce 1992,
Dougherty et al. 1995).

It is still somewhat puzzling that among the bright LMXB containing
neutron stars there is only one case with a jet (Cir X-1). Also, it is not
clear why jets have been found only in a few of the relatively frequent (1-2
yr$^{-1}$) bright X-ray transients. These systems are believed to be all
rather similar. This  indicates that there may be additional factors
determining the production of jets, factors for which no observational
counterparts have been identified so far.

Protostellar binaries are a bit of a different class of objects in this
context, since their disks are not fed by mass transfer. The effects of a
binary companion on a protostellar disk, however, may well be relevant for the
ability of the object to produce a jet. I return to this question in the last
section.

\section{Physics of magnetic acceleration: heuristics}
\index{jets!acceleration}\index{outflows!acceleration}
Magnetized winds and jets can be produced by rotating objects which, for one
reason or the other, have a magnetic field anchored in them. The importance of
such a magnetic field for the spindown of stars was realized by Schatzman
(1962) soon after the discovery of the solar wind. Quantitative models for
magnetized stellar winds were then developed by Weber and Davis (1967) and
Mestel (1968). The point of view in this work was, mostly, the spindown of
stars, but Michel (1969, 1973, see also Goldreich and Julian 1970) realized
the importance of the mechanism for producing high speed outflows, and
formulated relativistic models for magnetic
winds from pulsars. That strong outflows could also be driven magnetically by
accretion disks was proposed by Bisnovatyi-Kogan and Ruzmaikin (1976),
Blandford (1976) and Lovelace (1976). Workable quantitative models for such
flows were first produced by Blandford and Payne (1982), while full numerical
solutions for the steady (nonrelativistic) problem were first obtained by
Sakurai (1985, 1987).

\begin{figure}[tp]
\mbox{}\hfill\epsfysize6.5cm\epsfbox{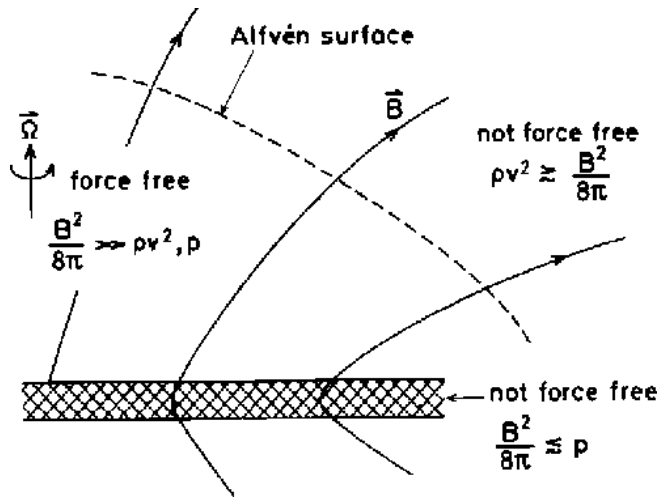}\hfill\mbox{}
\caption{\label{hsregs}
Regions of force-free and non-force free magnetic field in a disk-driven
wind.}
\end{figure}

In this section I review, at a heuristic level, the basic ideas and processes
involved. The theory is introduced a bit more formally in the next section.
For definiteness a wind generated by an accretion disk is considered here.
Most of the discussion however, applies equally well to rotating stars.

First, divide the disk and its surrounding space according to the relative
importance of the magnetic field energy density, see \fig{hsregs}. The
magnetic field to be used for producing the wind is anchored in the disk, so
its energy density there must be less than the rotational kinetic energy in
the disk. It can exceed the thermal energy density in principle, but in any
case its strength and configuration is determined by other forces, which
provide the anchoring of the field. Just how large the field strength is, for
any kind of observed disk, is still unknown since no field strengths have been
measured yet\footnote{An exception is the field strength (about 1 G) in the
protosolar nebula inferred from meteorites (e.g.\ Cisowski and Hood 1991)}.
Theoretical arguments allow for field strengths of the order of equipartition
with the gas pressure, in the case of dynamo-generated\index{dynamos} fields
(e.g.\ Hawley et al.\ 1995, Brandenburg et al.\ 1995), or even larger field
strengths for magnetic
flux dragged in with the accretion flow (Spruit, Stehle and Papaloizou 1995,
Lubow and Spruit 1995). Without being too specific, I assume that the
vertical field strength, at the disk surface, is reasonably large, since
the magnetic acceleration mechanism requires a field of some strength (a more
specific criterion is given in \sect{hscwd}). Outside
the disk, the gas density is typically so low (assuming a cool disk) that the
magnetic energy density is large compared with the thermal and rotational
energies. The field in this region must therefore be {\em force free}, much
like the solar corona (e.g.\ Foukal 1990). In the absence of torques acting
on
it, it must even be a {\em potential field}. As \fig{hsregs} suggests, we
are assuming that the field is of uniform polarity, over the region of the
disk where the wind is generated. Loops of field connecting different parts of
the disk cannot be excluded a priori. Such loops are sheared rapidly by
differential rotation, giving rise to a rich and not very well understood
complex of phenomena which is outside the scope of this discussion. It
suffices that a certain minimal fraction of the disk's magnetic flux does not
loop back to the disk surface, but is open to infinity.

As the flow is accelerated, the field strength drops due to the increasing
distance from the disk. The acceleration effectively stops when the flow speed
reaches the local Alfv\'en speed in the flow. The place where this happens is
called the Alfv\'en surface. Thus, outside the region where the magnetic field
dominates there is again a region where the field is not force free; in this
case because of inertial forces (\fig{hsregs}).

\begin{figure}[tp]
\mbox{}\hfill\epsfysize4.7cm\epsfbox{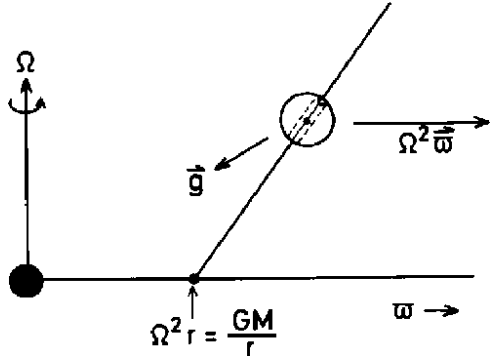}\hfill\mbox{}
%\picplace{4.7 cm}
\caption{\label{hsbead}
Bead-on-a-wire analogy for centrifugal acceleration by a magnetic field.}
\end{figure}

The acceleration process is illustrated in \fig{hsbead}. Assume that the
disk is cool, so that its rotation is close to Keplerian, and its thickness
can be neglected. Also assume that the gas is sufficiently ionized everywhere
that ideal MHD can be used, i.e. that gas is tied to the field lines. These
assumptions are not essential and can be relaxed in numerical models, such as
those of K\"onigl (1989). Assuming we are close to the disk surface, and the
field strength large, the atmosphere of the disk is forced to corotate with
the field lines sticking out of the surface. Since the Lorentz force only has
components perpendicular to the field, the gas is free to move along the field
line, under the influence of the other forces, like a `bead on a wire'.
At the foot point of the field line, the inward force of gravity just balances
the centrifugal force, because of our assumption of Keplerian rotation in the
disk. Along the field line, the centrifugal force increases with distance from
the axis. When the component of the centrifugal force along the field line
exceeds that of gravity, the gas tied to the field line is accelerated
outward.

This centrifugal process stops when the flow speed becomes comparable to the
Alfv\'en speed; at that point, the field is no longer strong enough to enforce
corotation.

Depending on one's preferences, the acceleration can also be
described in purely magnetic terminology (e.g.\ Lovelace, Wang and Sulkanen
1987), without appealing to a centrifugal force. This is discussed in
\sect{hscorm}.

\begin{figure}[tp]
\mbox{}\hfill\epsfysize 6cm\epsfbox{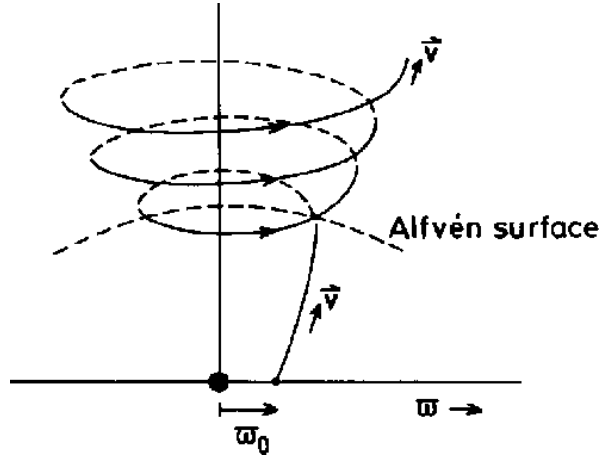}\hfill\mbox{}
%\picplace{6 cm}
\caption{\label{hstwist}
Development of the azimuthal field. With each rotation of the field line a
loop of field is added to the flow at the Alfv\'en surface.}
\end{figure}

Beyond the Alfv\'en surface, the inertia of the gas causes it to lag behind
the rotation of the field line, so that the field gets wound up. The simplest
way
to visualize this is by ignoring the rotation of the gas altogether. Then for
each rotation of the foot point of the field line, one loop of field is added
at the Alfv\'en surface. As the flow carries these loops away, a spiral shaped
field formed (\fig{hstwist}) with pitch $v/\Omega$, where $v$ is the flow
speed and $\Omega$ the rotation rate of the foot point. In a (nonrotating)
frame comoving with the flow, one sees a nearly azimuthal magnetic field, its
strength decreasing with time. The curvature force in the azimuthal
field is directed towards the axis, causing the flow to `collimate', i.e. to
become parallel to the rotation axis\index{outflows!collimation}.

\begin{figure}[tp]
\mbox{}\hfill\epsfysize=10cm\epsfbox{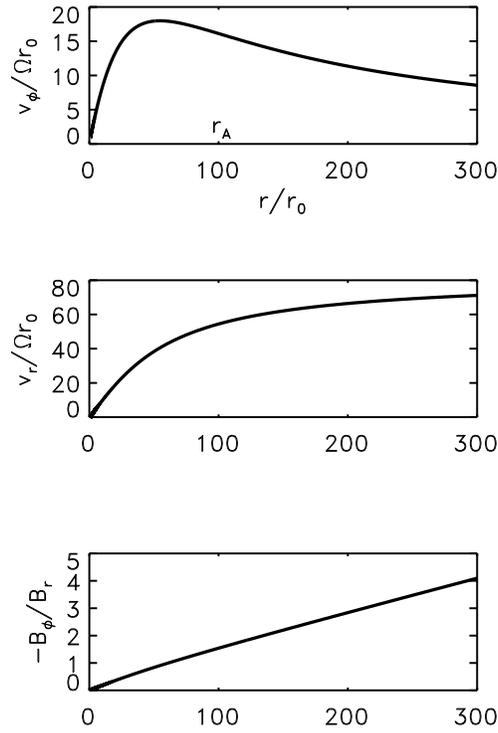}\hfill\mbox{}
\caption{\label{hsvrvf} Example of a magnetic wind model.
Top 2 panels: rotation velocity (measured in an inertial frame) and radial
velocity in units of the rotation speed at the base of the flow, as functions
of distance. Alfv\'en radius is at $100 r_0$. Lower panel: field angle.
(Cold WD model for $\mu=10^{-6}$, see text).}
\end{figure}

A numerical example of a magnetically accelerated flow is shown in
\fig{hsvrvf}. Note that the rotation velocity peaks near the Alfv\'en radius
\index{Alfv\'en radius} $r_{\rm A}$, which is at 100 times the starting
distance $r_0$ in this
example. At large distance, the rotation drops roughly in accordance with
angular momentum conservation. The asymptotic radial velocity is larger than
the rotation speed at $r_{\rm A}$ by a factor of order unity.

One of the attractions of the magnetic wind model is that it not only produces
outflow, but in principle can also take out the angular
momentum\index{disks!angular momentum loss} from the
disk, allowing it to accrete (Blandford 1976, Bisnovatyi-Kogan and Ruzmaikin
1976). To see how effective this can be let us estimate the angular momentum
flux carried by the wind. Since the flow corotates roughly up to the Alfv\'en
point, the specific angular momentum carried is of the order $\Omega r_{\rm
A}^2$, hence
\be \dot J_{\rm w}=\dot M_{\rm w}\Omega r^2_{\rm A} ,\label{hsjdot}\ee
where $\dot M_{\rm w}$ is the mass flux on a field line with foot point $r_0$.
It turns out that this estimate is
actually exact (\sect{hsstead}). The angular momentum that has to be
extracted (locally at $r_0$) from a Keplerian disk in order for it to accrete
at a rate $\dot
M_{\rm a}$ is $\dot J_{\rm a}={1\o 2}\Omega r_0^2\dot M_{\rm a}$, hence
\be
{\dot M_{\rm w}\o\dot M_{\rm a}}={1\o 2}\left({r_0\o r_{\rm A}}\right)^2.
\label{hsmm}
\ee
Since this relation is exact, and $r_{\rm A}$ always larger than $r_0$, it
follows that only a fraction of the mass flux in the disk can flow out in the
wind. It is possible in principle, however, that the wind carries away all
angular momentum that the disk has to loose in order to accrete, i.e. without
angular momentum transport by viscous torques in the disk. Such a disk would,
in the absence of viscous dissipation, be silent. This is verified by
looking at the energy balance. The energy flux $\dot E_{\rm w}$ in the wind is
given by $\Omega\dot J_{\rm w}$ (the work done against the wind torque). If
all the angular momentum is carried with the wind we have $\dot E_{\rm w}=\dot
E_{\rm a}$ (using \eq{hsmm}), where $\dot E_{\rm a}=1/2(\Omega
r_0)^2\dot M_{\rm a}$ is the rate of gravitational energy release in a
Keplerian disk. Thus, if the wind carries away all the angular momentum, it
also carries away all the accretion energy.

\begin{figure}[tp]
\mbox{}\hfill\epsfysize 4cm\epsfbox{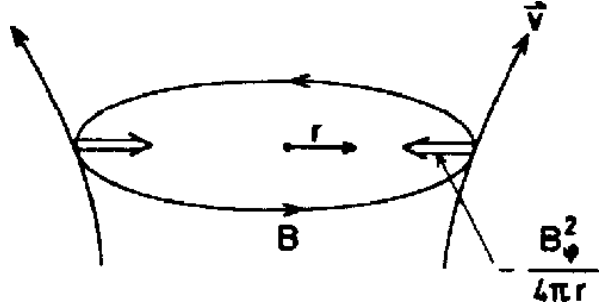}\hfill\mbox{}
%\picplace{4 cm}
\caption{\label{hshoop}
Collimation of the flow by the curvature force of the azimuthal magnetic
field.}
\end{figure}

\subsection{Structure of a disk driven wind}
The magnetically driven wind is conveniently broken down
into three conceptual stages or subprocesses. Near the disk surface, the
wind is `launched': details of disk structure and magnetic field configuration
near this surface determine how much mass is launched into a flow. After this,
the flow can be regarded as ballistic, its acceleration being governed almost
entirely by gravitational and magnetic/centrifugal forces. After the
acceleration phase, which ends roughly at Alfv\'en surface, the collimation
phase starts, in which the flow is deflected towards the axis by `hoop stress'
(\fig{hshoop}).

Since the outer radius of a disk is typically much larger than the inner
radius, conditions can vary dramatically with distance in the disk. The wind
problem is, therefore, a function of distance. In the inner regions where the
field strength and the rotation speed are large, conditions are favorable for
producing high speed collimated winds. In the outer regions, one would expect
lower wind speeds, and probably less collimation. The fluxes of mass and
angular momentum from the outer regions, on the other hand, could be large
compared to flows from the inner regions. Flows from these regions may well
coexist. Since the magnetic field inside the Alfv\'en radius is force free,
its configuration is determined by a global force balance. The wind properties
from adjacent regions in the disk are therefore coupled somewhat, through
their dependence on the shape of the poloidal field. This aspect of the disk
wind problem has not yet received much attention.

In the next sections this picture is elaborated a bit more formally. In the
process, some problem areas of current interest will be noted, relating, in
particular, to the launching and collimation phases.

\section{Steady axisymmetric magnetic flows}
\label{hstheor}
The theory of magnetically driven flows from  rotating objects has been
given in many texts (e.g.\ Weber and Davis 1967, Mestel 1968, Heinemann and
Olbert 1978, Okamoto 1975, Blandford and Payne 1982).
I repeat the basic derivation here, under the assumption of ideal
magnetohydrodynamics, in the nonrelativistic limit. The extension to
relativistic MHD can be found in Michel (1969), Goldreich and Julian (1970),
Bekenstein and Oron (1978), Camenzind (1987), for a general treatment of
relativistic MHD see Lichn\'erowicz (1967).

\label{hsstead}
The basic assumptions made are the MHD approximation, and that the flow is
stationary and axisymmetry. The MHD approximation is justified if the rotating
object actually produces an outflow of any observational significance; it
holds if the density of charge carriers is large compared with the so-called
Goldreich-Julian\index{Goldreich-Julian density} density, $N_{GJ}=\Omega
B/(4\pi c e)\sim 10^{-2} \Omega B$
cm$^{-3}$. This limit is of importance in the case of pulsar magnetospheres
(Goldreich and Julian 1969), but is so low that it is not likely to become
relevant for most of
the observable winds and jets. Though deviations from stationarity are implied
by the production of outward traveling `knots' in most jets (for an example
see Mirabel and Rodriguez 1994), the time scale for acceleration is still
likely to be short compared to the time scale of these variations, so that
stationarity is a good assumption during the acceleration phase. Much of the
jet phenomenology is consistent with axisymmetry. Theoretically,
nonaxisymmetric instabilities are likely to become important in parts of the
jet; this is discussed further below. Keeping this in mind, we proceed with
the axisymmetric case. The final assumption made is that of infinite
conductivity. This can easily be relaxed (K\"onigl 1989), but deviations from
this approximation are important only in the case of protostellar jets, where
the flows are so cool that the conductivity needs to be considered in detail.

\begin{figure}[tp]
\mbox{}\hfill\epsfysize5.6cm\epsfbox{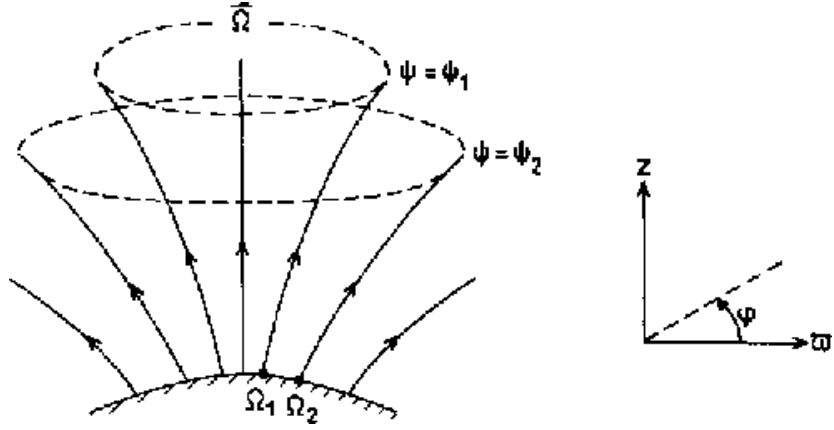}\hfill\mbox{}
%\picplace{5.6 cm}
\caption{\label{hsaxi}
Axisymmetric poloidal magnetic anchored in a rotating object. Magnetic
surfaces are labeled by the flux function $\psi$. Rotation rate can be a
function of $\psi$.}
\end{figure}

The equations for stationary ideal MHD are (e.g.\ Roberts, 1967)
\be \nabla\times({\bf v}\times{\bf B})=0, \label{hsind}\ee
\be
\rho {\bf v}\cdot\nabla{\bf v}=-\nabla p -\rho \nabla\Phi+{1\over 4\pi}
(\nabla\times{\bf B})\times{\bf B}, \label{hsmot}
\ee
\be \nabla\cdot(\rho{\bf v})=0, \label{hscon}\ee
\be \nabla\cdot{\bf B}=0.\label{hsdiv}\ee
Here ${\bf B,v},\Phi,p,\rho$ are the magnetic field vector, the velocity, the
gravitational potential, the gas pressure and the density.
Using cylindrical coordinates ($\po,\phi,z$), axisymmetric vectors like
$\bf B$ and $\bf v$ can be written in terms of their poloidal ($_\rp$)
and toroidal ($_\phi$) components:
\be {\bf B}={\bf B}_\rp+B_\phi{\bf e}_{\phi}, \label{hsbpt}\qquad
{\bf v}={\bf v}_\rp+v_\phi{\bf e}_{\phi}, \ee
where the poloidal components lie in the meridional ($\po,z$) plane.

With the assumed axisymmetry, these equations have a great deal of symmetry,
so that they can, in part, be reduced to algebraic equations. This is done as
follows. Due to axysimmetry and div${\bf B}=0$, $\bf B_\rp$ can be written
in terms of a {\em flux function} $\psi$:
\be {\bf B}_\rp={1\over\po}\nabla\psi\times{\bf e}_\phi, \label{hsipt}\ee
i.e. $B_z=1/\po\,\p\psi/\p \po$, $B_r=-1/\po\,\p\psi/\p z$. The flux function
(also called stream function or vector potential) is constant along field
lines:
\be {\bf B}\cdot\nabla\psi=0 .\ee
Thus, $\psi$ can be read as a {\em label} numbering the field lines of the
poloidal field (\fig{hsaxi}). It plays an important role in this numbering
capacity, since
there will be several scalar fields $\alpha$ with the property
$\bp\cdot\nabla\alpha=0$. Such fields have their gradients parallel to that of
$\psi$, and therefore are functions of $\psi$ only, and are also constant on
field lines.
From (\eq{hsind}) we have ${\bf B}\times{\bf v}=\nabla f$, where $f$ is an
axisymmetric scalar. Writing this out into poloidal and toroidal components:
\be \vp\times\bp+v_\phi\efi\times\bp+B_\phi\vp\times\efi=\nabla f.
\label{hscross}\ee
The toroidal component of this is
\be \vp\times\bp=0,\qquad{\rm or}\qquad \vp=\kappa(\po,z)\bp. \label{hspar}
\ee
Thus, the poloidal velocity is parallel to the poloidal magnetic field, a
consequence of the infinite conductivity assumed. With (\ref{hspar}), the dot
product of (\ref{hscross}) with $\bp$ yields $\bp\cdot\nabla f=0$, so that $f$
is not just a scalar, but a function of the field line number $\psi$ only.
With this, (\ref{hscross}) yields
\be v_\phi-\kappa B_\phi=\po f^\pr(\psi), \label{hsom0}\ee
where $f^\pr=\rd f/\rd\psi$. With (\ref{hspar}), (\ref{hsdiv}), the continuity
equation (\ref{hscon}) yields $\bp\cdot\nabla(\rho\kappa)=0$, so that
\be \rho v_\rp/B_\rp=\rho\kappa=\eta(\psi), \label{hseta} \ee
for some function $\eta$. This equation simply states that the mass flux
density, per unit of poloidal magnetic flux, is constant along a field line:
each field line has its own mass flux, in this sense.

Now find the point along a poloidal field line inside the rotating object, on
which $B_\phi=0$, and call this the `foot point' of the field line. If the
object, together with $\bp$, is symmetric about the equator, this point is on
the equatorial plane\footnote{Such symmetry simplifies the visualization, but
is not necessary. All poloidal field lines are `bent back' at the object's
surface by the wind torque in the same direction (lagging behind the
rotation), so that $B_\phi$ is of opposite signs at two successive crossings
of the field line through the object's surface. Hence on all poloidal field
lines crossing the surface the toroidal component changes sign at some point
inside the object.}. Let
$\Omega=v_{\phi 0}/\po_0$ be the rotation rate of this foot point. Then we
find from (\ref{hsom0}) that $f^\pr(\psi)=\Omega$.
Loosely, we can call this the
`rotation rate of the field line'. Remember, however, that the plasma rotation
rate, $\Omega_\rp=v_\phi/\po$, is {\em not} constant along a field line. It
cannot be, because the plasma corotates with the foot point only as
long as the field is strong enough to dominate over the plasma; the rotation
starts lagging behind at larger distances from the object where the field is
weaker. Let ${\bf v}^\pr$ be the flow velocity measured in a frame rotating
with the angular frequency $\Omega$:
\be{\bf v}^\pr={\bf v}-\po\Omega(\psi)\efi \label{hsvpr}\ee
Then (\ref{hsom0}) and (\ref{hspar}) can be combined into
\be {\bf v}^\pr=\kappa{\bf B}.\label{hsbpar}\ee
In other words, in a frame corotating with the foot point of a field line, the
flow is everywhere parallel to the magnetic field (this frame can be different
for each poloidal field line, if the object rotates differentially). Whereas
the {\em poloidal} flow component is parallel to the {\em poloidal} field
component in both the rotating and the stationary frames, the total velocity
is parallel to the total $B$-field only in a corotating frame.

Next, consider the equation of motion (in a non-rotating frame). Using
axisymmetry and the identity
$(\nabla\times{\bf B})\times{\bf B}=-\nabla B^2/2+({\bf B}\cdot\nabla) {\bf
B}$, the toroidal component is
\be \rho({\bf v}\cdot\nabla{\bf v})_\phi={1\over 4\pi}({\bf B}\cdot\nabla{\bf
B})_\phi. \ee
With the vector relation $({\bf a}\cdot\nabla{\bf b})_\phi={\bf a}\cdot \nabla
(\po b_\phi)/\po$, (\ref{hspar}) and (\ref{hseta}), this can be written as
\be
\bp\cdot\nabla(\rho\kappa\po v_\phi)={1\over 4\pi}\bp\cdot\nabla(\po
B_\phi).
\ee
This can be integrated, yielding
\be
{1\o B_\rp}(\rho v_\rp\po v_\phi-{\po\over 4\pi}B_\phi B_\rp)
=\eta L, \label{hshp}
\ee
where $L$ is a a function of $\psi$ only.
Hence
\be \po(v_\phi-{1\over 4\pi\eta}B_\phi)=L(\psi). \label{hsh}\ee
The first term in (\ref{hshp}) is the flux of angular
momentum by the poloidal flow, the second the magnetic torque. Thus,
(\ref{hshp}) expresses that the total flux of angular momentum, per unit of
poloidal magnetic flux, is constant along each field line. The total angular
momentum flux measured per unit of {\em mass} flowing along the field line is
$L$.

Before proceeding with the equation of motion, we use
(\ref{hsom0}) to eliminate $B_\phi$ from (\ref{hsh}). This yields
\be
v_\phi-\Omega\po={L-\Omega\po^2\o\po[1-1/(4\pi\kappa^2\rho)]}. \label{hssin}
\ee
The denominator on the RHS vanishes when $4\pi\rho{v_\rp^2/ B_{\rm
p}^2}=1$, or $v_\rp=v_{\rm Ap}$, where $v_{\rm Ap}=B_{\rm
p}/(4\pi\rho)^{1/2}$ is the poloidal Alfv\'en speed. The location along the
field line where this happens is called the {\em Alfv\'en
point}\index{Alfv\'en radius}, because an
axisymmetric Alfv\'en wave propagates along a magnetic surface at the speed
$v_{\rm Ap}$ (independent of the value of $B_\phi$). If the flow
is to be accelerated to values beyond the local poloidal Alfv\'en speed, the
numerator in (\ref{hssin}) also has to vanish at the Alfv\'en point. If we
denote the physical variables at the Alfv\'en point with a subscript $_{\rm
A}$, this condition yields
\be L=\Omega\po_{\rm A}^2. \label{hspia}\ee
The interpretation of the singularity at $\ra$ is similar to that of the sonic
point in the transition from subsonic to supersonic flow in ordinary
hydrodynamics (jet nozzle, stellar winds). Eq. (\ref{hspia}) has an important
physical interpretation.

Eq. (\ref{hspia}) has an important
physical interpretation. If the flow were to corotate with the field up to
$\po_{\rm A}$ (which actually it does not, see above) it would, at that point,
have the specific angular momentum given by (\ref{hspia}). Since $L$ is the
specific angular momentum flux in the wind (including the magnetic torque!),
it is as if the flow were kept rigidly corotating out to $\po_{\rm A}$, and
then released without further magnetic torques. As far as the angular momentum
flux is concerned, the flow `effectively corotates' out to $\po_A$.

As in ordinary dissipationless hydrodynamics, the equation of motion in the
direction of the flow can be integrated in terms of a Bernoulli function,
provided that the energy equation is sufficiently `simple'. In practice this
is the case if polytropic or isothermal equations of state can be used as
approximations:
\be P=K\rho^\gamma, \ee
where $K,\gamma$ are constants. The isothermal case is obtained with
$\gamma=1$. These are not necessarily very good approximations to the real
situation, since energy dissipation and cooling processes usually are present.
These play an important role in the early stages of the acceleration, see
\sect{hslaunch}. Once the flow has been accelerated beyond the speed of
sound, however, the dynamics does not depend much on the temperature of the
gas any more.

In a corotating frame (corotating with the foot point of the field line, cf.
the discussion above), the equation of motion becomes:
\be
\rho{\bf v}^\pr\cdot\nabla{\bf v}^\pr=-\nabla p +
{1\over 4\pi}(\nabla\times{\bf B})\times{\bf B}-\rho\nabla\Phi+
\rho\Omega^2{\bfg{$\po$}}+2\rho{\bf v^\pr}\times{\bf \Omega}, \label{hsreqm}
\ee
where $\bfg{$\po$}=\po{\bf e}_\phi$ and $\Phi$ is the gravitational potential:
\be \Phi=-{GM\over r} .\ee
Let ${\bf s}$ be a unit vector parallel to the magnetic field. Taking the dot
product of (\ref{hsreqm}) with $\bf s$, we get the component of the equation
of
motion parallel to the field lines. I denote the derivative along the field
line by
\be \p_s \equiv {\bf s}\cdot\nabla . \ee
Then the centrifugal term $\Omega^2{\bf s}\cdot{\bfg{$\po$}}$ can be written
as $\p_s(\Omega^2\po^2/2)$, since $\Omega$ is a constant along a field line
(remember it is the foot point rotation rate, not the local fluid rotation).
The Coriolis term disappears because it is perpendicular to $\bf v$, which is
parallel to $\bf B$ in the rotating frame. The magnetic term disappears
because the Lorentz force is perpendicular to $\bf B$. Thus (\ref{hsreqm})
yields
\be
\p_s{1\over 2}v^{\pr 2}=-{1\over \rho}\p_s p -\p_s\Phi +\p_s({1\over
2}\Omega^2\po^2).
\ee
With the polytropic equation of state, the thermal term can be written as
\be {1\over \rho}\p_s p=\p_s(k c_\rs^2), \ee
where $k$ is a factor of order unity:
\bea
k&=&{\gamma\over\gamma -1} \qquad (\gamma \neq 1),\\
 &=& \ln(\rho) \qquad  (\gamma = 1),\nonumber
\eea
and
\be c_s=(p/\rho)^{1/2} \ee
is the isothermal sound speed. The equation can now be integrated; with the
definition $v^\pr$, this yields:
\be
{1\over 2}v_\rp^2+{1\over 2}(v_\phi-\Omega\po)^2+k c_\rs^2+\Phi-{1\over
2}\Omega^2\po^2=E(\psi), \label{hsbern}
\ee
alternatively:
\be
{1\over 2}v^2-v_\phi\Omega\po+k c_\rs^2+\Phi=E(\psi). \label{hsberna}
\ee
The integration constant $E$ depends on the field line label $\psi$ only.
This is called the {\em Bernoulli equation}. It states that, in the rotating
frame, the sum of kinetic, thermal, gravitational and a `centrifugal energy'
is constant along a field line. A look at the terms in this equation shows the
basics of centrifugal acceleration. Assume that the field is strong enough to
enforce approximate corotation, so that $v_\phi\approx\Omega\po$. Assume
that we are looking at a flow which has already been accelerated to supersonic
speeds, $v_\rp\gg c$, so that the thermal term can be ignored compared with
the first term. The kinetic energy ${1\o 2}v_\rp^2$ then increases with
distance $\po$ from the axis due to the rapid decrease of the centrifugal
term. This is offset by the increase of the gravitational potential $\Phi$,
but eventually the centrifugal term dominates because $\Phi$ reaches a
constant value at
infinity. More precisely, the condition for outward acceleration, at any point
on the field line, and still ignoring thermal effects, is:
\be \p_s(\Phi-{1\over 2}\Omega^2\po^2) < 0 \qquad(T=0),\label{hsaco}\ee
where the arc length $s$ is taken to increase in the direction of $\po$.
The thermal term adds to the acceleration: as the flow expands outward,
thermal energy is converted into kinetic energy, in the same way as in a jet
nozzle. In slowly rotating stars, this is the dominant process driving the
stellar wind, and one has a {\em thermally driven} wind\index{stellar winds}
(Parker 1963, for an introduction see Foukal 1990). In
more rapidly rotating stars, and in our case of jets produced by disks, the
thermal energy plays a role only in the initial stages, and most of the actual
acceleration is centrifugal. The conditions for launching a wind from a disk
are discussed further in \sect{hslaunch}.

\subsection{Acceleration: centrifugal or magnetic?}
\index{jets!acceleration}\index{outflows!acceleration}
\label{hscorm}
This discussion above suggests that the centrifugal force could accelerate the
flow
indefinitely, but this is an artefact of our assumption of corotation. When
the field becomes weak with distance, the azimuthal velocity starts lagging
behind $\Omega\po$, and then becomes small compared with $\Omega\po$.
The second term in \eq{hsbern} then nearly cancels the centrifugal term. How
much acceleration still remains depends on the details of how fast $v_\phi$
decreases, and the other equations have to be used as well to determine this.

By working in the rotating frame there is no contribution from magnetic
forces. This may seem strange, since it is ultimately the magnetic forces that
transmit the rotational energy of the object in which they are anchored to the
flow. That the acceleration can also be regarded as magnetic is seen by
considering the equation of motion in an inertial frame. We want to know how
the poloidal velocity is accelerated by the magnetic field. The poloidal
component of the Lorentz force is
\be {\bf F}_\rp={1\over 4\pi}[(\nabla\times{\bf B})\times{\bf B}]_\rp=
{1\over 4\pi}[(\nabla\times{\bf B})_\rp\times{\bf B}_\phi+(\nabla\times{\bf
B})_\phi\times{\bf B}_\rp], \label{hsfl}
\ee
where $_\rp$ and $_\phi$ are the poloidal and toroidal components of vectors,
as
defined above.
The second term is perpendicular to ${\bf B}_\rp$ and ${\bf v}_\rp$, so does
not contribute to acceleration. Thus the accelerating force is the first term
on the right in (\ref{hsfl}), which can be written as
\be {1\over 4\pi}(\nabla\times{\bf B}_\phi)\times{\bf B}_\phi. \ee
The condition that the poloidal flow is accelerated is then $\vp\cdot
(\nabla\times{\bf B}_\phi)\times{\bf B}_\phi>0$. With (\ref{hsbpt}) this can
be written as:
\be
-\vp\cdot[\nabla{B_\phi^2\over 8\pi}+{B_\phi^2\over 4\pi}{\bf e}_\po]>0.
\label{hsmac}
\ee
This shows that it is the pressure gradient (first term) and tension force
(second term) of the {\em toroidal} field which determine the acceleration.
For a net outward acceleration to occur, $B_\phi^2$ has to decrease outward
along the field line sufficiently rapidly to overcome the tension force, which
is directed towards the axis. Whether this is actually the case can not be
determined from this argument, since one has to solve the full problem to find
$B_\phi$.

We have derived both acceleration conditions (\ref{hsaco}) and (\ref{hsmac})
from the same equation of motion, hence the {\em magnetic} and {\em
centrifugal} points of view are equivalent. This can be verified by deriving
the Bernoulli equation in an inertial frame. The component of the equation of
motion
parallel to $\bf v$ then has a magnetic term ${\bf v}\cdot(\nabla\times{\bf
B})\times{\bf B}/(4\pi)$ instead of the centrifugal term. Using (\ref{hseta}),
(\ref{hspar}), (\ref{hshp}), the magnetic fields in this term can be replaced
by velocities. The end result is \eq{hsberna}.

The situation at hand determines which of
these views is more appropriate. In regions where the field is strong enough
to enforce approximate corotation, the centrifugal view is appropriate. When
corotation is not a good
approximation, the acceleration is more conveniently viewed as due to the
magnetic pressure of the azimuthal field. Corotation is usually a good
approximation up to the Alfv\'en radius (with a significant exception, see
\sect{hscwd} below). Beyond the Alfv\'en radius, the field lines stop
corotating, and instead are rapidly wound up into a nearly toroidal field.
Here, some residual acceleration by the gradient of $B_\phi^2$ takes place.

\section{Acceleration in a fixed poloidal field }
To the extent that the field above the disk can be approximated by a potential
field, it depends only on the distribution of its `sources' on the disk,
namely the normal field component $B_z(\po,z=0)$. The accelerating flow and
the toroidal field which develops in it, however, exert forces which distort
the poloidal field. A full solution of the problem therefore requires solving
the equation of motion in the direction perperdicular to the poloidal field
lines, the `cross-field' equation. This is considered below (section
\ref{hscro}). A convenient
approximation for the wind problem, however, is to consider the {\em poloidal}
field as fixed and given. The toroidal field (which is responsible for the
acceleration), is left free, to be solved for. This approximation is good for
dealing with the launching and acceleration aspects of the problem in the case
when the Alfv\'en radius is at a large distance, since most of the
acceleration then takes place in the magnetically dominated region. It
obviously breaks down whereever the toroidal field dominates over the poloidal
component. Thus collimation of the flow by the `hoop stress' of $B_\phi$
cannot be dealt with in this approximation. It also fails, in the entire
domain, in the high-mass loss regime discussed in \sect{hscwd}.

Having dispensed with the cross-field equation by the fixed-poloidal-field
approximation, the flow can be solved on each poloidal field line separately.
The solution of this problem is determined by eqs. (\ref{hsbern}),
(\ref{hssin}),
(\ref{hseta}) above. On each of the
field lines, $\eta$, $\ra$ and $E$ are constants, still to be found from the
solution. It turns out (Sakurai, 1985) that the problem is visualized
conveniently in a space in which the coordinates are $s$, the arc length along
a poloidal field line, and $\rho$, the gas density. The Bernoulli equation can
then be read as an algebraic equation specifying  a relation beteen $\rho$ and
$s$. This relation is the solution of the problem. To see this, note that for
each field line, $B_\rp$ is a known function of $s$, and $\po$ is a known
function of $s$ through the known shape of the poloidal field lines. Hence
with (\ref{hseta}) the first term in (\ref{hsbern}) is of the form
$v_\rp^2/2=f_1(s,\rho;\eta)$. In this notation, the semicolon separates the
the coordinates $s,\rho$ from the parameters $\eta,E,\ra$. With the rotation
rate $\Omega$ of each field line
specified, eqs. (\ref{hssin}),(\ref{hspia}) show that the second term is of
the
form $v_\phi^{\pr 2}/2=f_2(s,\rho;\ra)$. The gravitational and centrifugal
terms are functions of $s$ only, and the thermal term is of the form
$f_3(\rho;K)$ in the polytropic case, or $f_3(\rho;c_\rs)$ in the isothermal
case. The conditions at the surface of the disk have to be known to solve the
problem, so we may assume that the values $\rho_0,p_0$ of the pressure and
density  at some point near the base ($s=0$) of the flow are known. One of
these 2 values then determines $K$ or $c_\rs$ directly while the other,
$\rho_0$ say, fixes (through the solution of the problem) a relation between
the unknown constants ($\eta,\ra,E$). Thus, for a solution of the problem, two
more conditions are needed to specify all three constants. These are
two critical point conditions, which appear as follows.

\subsection{Critical points}
\index{critical points}
Writing the Bernoulli equation in the form
\be H(s,\rho;\eta,\ra)=E, \ee
the solution curve $\rho(s)$ can be regarded as a contour line of the
Bernoulli function $H$ in the $s,\rho$ plane. The astrophysically relevant
solutions start at a very high density near the disk surface, and decrease to
vanishing
density at infinity. We are therefore interested in unbroken contours of $H$
which cover the entire $\rho$ coordinate. $H$ has `mountain ranges' however,
and brief inspection will show that these mountain ranges can be crossed by a
level countour only through mountain passes. These mountain passes are
critical points of the saddle type. As we shall see, there are two of these
points, and the solution has to cross both. The elevation of $H$ at one of
the points determines the value of $E$. For the solution also to cross the
other point, this point must have the same elevation. This will in general be
the case only for certain combinations of the parameters $\eta,\ra$. Together
with the given value of $\rho_0$, the two critical points thus determine the
unknown parameters $\eta$, $\ra$ and $E$ and the problem is solved.

Notice that there is no need to solve any differential equations, the whole
magnetic wind problem is algebraic, for any configuration of the poloidal
field (as long as it is prescribed in advance).

It remains to be shown that there are two relevant critical points. At a
critical point we have
\be \p H/\p s=0,\quad \p H/\p \rho =0. \ee
Substituting $v_\rp$ and $v_{\rm p\phi}$ from (\ref{hseta}) and (\ref{hssin})
one finds that
\be
\rho{\p H\over\p\rho}=-{v_\rp^4-v_\rp^2( c_\rs^2+v_{\rm Ap}^2+V_{\rm A\phi}^2)
+ c_\rs^2v_{\rm Ap}^2 \o v_\rp^2-v_{\rm Ap}^2}, \label{hscpt}
\ee
where $v_{\rm Ap}$, $v_{\rm A\phi}$ are the Alfv\'en speeds based on the
poloidal and toroidal field strengths, respectively:
\be
v_{\rm Ap}={B_\rp^2\over 4\pi\rho}, \quad
v_{\rm A\phi}={B_\phi^2\over 4\pi\rho}.
\ee
To interpret the expression for $\p H/\p\rho$, note that the dispersion
relation for magnetosonic waves\index{magnetosonic waves} in
a homogeneous medium is (e.g.\ Plumpton and Ferraro, 1966):
\be u^4-u^2( c_\rs^2+v_{\rm A}^2)+ c_\rs^2v_{\rm A}^2\cos^2\theta=0,
\label{hsms}\ee
where
\be u=\omega/k \ee
is the wave speed\footnote{magnetosonic waves, though anisotropic, are
nondispersive.}, and $\theta$ the angle between the magnetic field vector
and the direction of the wave vector $\bf k$. Thus, $\p H/\p \rho=0$ when the
poloidal velocity equals the speed of a magnetosonic wave propagating
parallel to the poloidal flow (so that $\cos\theta=B_\rp/B$). Thus
(\ref{hscpt}) can be written as:
\be
\rho{\p H\over\p\rho}= - {(v_\rp^2-v_{\rm sp}^2)(v_\rp^2-v_{\rm
fp}^2)\over v_\rp^2-v_{\rm Ap}^2}, \label{hscp}
\ee
Where $v_{\rm sp}$ and $v_{\rm fp}$ are the solutions of (\ref{hsms}) for
$\cos\theta=B_\rp/B$. Critical points therefore occur when the flow just
balances the speed of a
magnetosonic wave propagating opposite to the flow. They are called the slow
mode critical point and fast mode critical point, or slow and fast point, for
short. In addition to these critical points, there is a singular point of a
different kind at $\vert v_\rp\vert=v_{\rm Ap}$. Note that only the slow and
fast mode critical points yield constraining
relations for the solution, the Alfv\'en point does not yield an additional
constraint. Through its appearance in the denominator in (\ref{hscp}), the
Alfv\'en point is a node rather than a saddle point; all solutions which pass
through the slow and fast points also pass through the Alfv\'en point. This is
because a critical point condition has already been applied at the Alfv\'en
point in deriving (\ref{hspia}). The Alfv\'en point, however, plays a new role
as a critical point when the cross-field balance is considered
(\sect{hscro}).

The practical problem of determining the location of the critical points and
computing the full solution depends a bit on the character of the poloidal
field specified. A simple case is the Weber and Davis (1967) model. This is
discussed further in \sect{hscwd} below.

\subsection{Multiple critical points}
Since the position of the critical points depends on the geometry of the
poloidal field configuration, one may wonder if there could not be more than
just 2 critical points. $H$ could have an additional mountain range such that
there are two slow points and a fast point, for example. Since each critical
point adds a condition that has to be satisfied by the flow, and
the problem is just closed with two critical point conditions, one might think
such multiplicity is excluded. The flow, however, has a way of generating
additional degrees of freedom. Except in carefully construed cases, a
supersonic flow develops a shock\index{shock waves} at a location where it is
forced to decelerate. Thus, if the shape of the poloidal field is sufficiently
complex that it forces the flow to decelerate somewhere, a shock is formed
near that location. This can happen if the flow diverges sufficiently rapidly
due to a decrease in poloidal field strength, or if the path of the field line
is such that it takes the flow up and down theeffective potential
(gravitational plus centrifugal) more than once. The Hugoniot conditions will
fix the properties of this shock, but the position of the shock is determined
only if one more condition is imposed. The flow can pass through several
critical points in such a way that each additional critical point is
associated with a shock. After each shock, the flow is reaccelerated and
passes though a new critical point. The regularity condition at the additional
critical point determines the position of the shock. There are two kinds of
shocks in magnetohydrodynamics, named slow and fast shocks (e.g.\ Jeffrey and
Taniuti, 1964), so that the formation of shocks can in principle take place in
association with additional slow critical points, as well as with additional
fast points. An example of an additional slow point in a realistic disk field
geometry is given in Cao and Spruit (1994). Cases with multiple fast points
are discussed in Heyvaerts and Norman (1989).

\subsection{Launching of the wind: the sonic point}
\label{hslaunch}
\index{outflows!launching}
If the flow is to be strongly accelerated, the Alfv\'en point must be at a
large distance, and this requires the field to be strong enough to enforce
corotation out to that large distance. In this case, it is a good
approximation to assume that the Alfv\'en speed is large compared to the sound
speed, at the base of the acceleration region. The slow mode speed, measured
at the slow mode critical point, is then close
to the sound speed, and the slow mode has the character of a
sound wave guided along the field line. For this reason, the slow point is
also called the sonic point. Without loss of physical generality, I equate the
slow mode speed to the sound speed for the rest of this section.

The importance of the sonic point for the wind problem is that it regulates
the {\em mass flux} on the field line; it governs how much mass is `launched'
into the accelerating region.
At the sonic point, $v_\rp=c_\rs$, and the mass flux (per unit area) is
\be \dot m=(\rho c_\rs)_{\rm c}, \ee
where the index $_{\rm c}$ means evaluation at the sonic point. The sound
speed
can be assumed to be known, either explicitly if an isothermal model is used,
in general by the energy balance model used. The mass flux is then known if
the density at the sonic point is known. At the sonic point, the pressure
balance along the flow is affected by hydrodynamic forces, but in the subsonic
region before the sonic point their influence is small. As a fair (order of
magnitude) approximation, we can take the pressure distribution to be
{\em hydrostatic} between the foot point (index $_0$) and the sonic point, and
supersonic beyond. An estimate of the mass flux is then obtained as
\be
\dot m\approx\rho_0 c_{\rm s} \exp[-(\Phi_{\rm ec}-\Phi_{\rm e0})/c_{\rm
sc}^2] \label{hsemd}
\ee
where the temperature has been approximated as constant, and $\Phi_{\rm e}$ is
the effective potential including the centrifugal term,
\be \Phi_{\rm e}(s)=-{GM\o r(s)}-{1\o 2}\Omega^2\po^2(s). \label{hsep}\ee
To complete the estimate, we need a value for the potential at the sonic
point. Assume that the potential has a maximum, measured along the
field line; first increasing due to the gravity of the central object, and
then decreasing due to the centrifugal force. Along the increasing part, the
density is stratified nearly hydrostatically. Approaching the maximum, the
mass starts flowing when the thermal energy in the gas becomes comparable to
the distance to the top of the potential barrier.
\begin{figure}[tp]
\mbox{}\hfill\epsfysize4.6cm\epsfbox{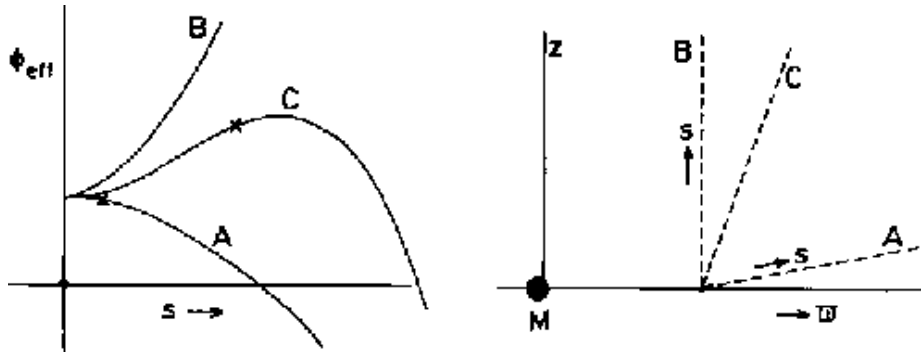}\hfill\mbox{}
%\picplace{4.6 cm}
\caption{\label{hsbarr}
Variation of the effective potential with arc length along field lines of
different inclinations.}
\end{figure}
At low temperatures, this happens close to the maximum of the potential. An
approximate mass flux is therefore found from (\ref{hsemd}) by taking for
$\Phi_{\rm ec}$ the maximum of $\Phi_{\rm e}$. For higher temperatures, the
sonic point occurs somewhat before the maximum of the potential. In the
absence of rotation the potential is due to gravity alone, and its maximum is
at infinity. In this case, the sonic point occurs roughly at the point where
the thermal energy is equal to the depth of the gravitational potential. Such
a flow is a thermally driven wind\index{stellar winds}, like the Sun's
(e.g.\ Foukal, 1990).
At low temperatures, the mass flux is very sensitive to the height of the
potential barrier, since it comes in exponentially.

For an understanding of the launching of the wind from a disk, we have to look
more closely at the variation of
the effective potential near the base of the flow. In \fig{hsbarr} the
variation of the potential is sketched for three paths starting at the same
point at the midplane of the disk. If the path is vertical (B), the potential
increases monotonically, there is no maximum sufficiently close to the disk
surface, and at best a feeble thermally driven wind is possible. At an
intermediate inclination (C), there is a maximum near the disk surface, and
conditions for launching a wind can be good, depending on the temperature of
the gas. Along a path close to the surface (A), the effective potential {\em
decreases monotonically}. In this case, the wind can start right from the disk
surface. The boundary between cases (C) and (A) occurs when the potential
curves neither up nor down at the foot point $s=0$ (see \fig{hsbarr}),
i.e. when
\be \p^2\Phi_e/\p s^2\vert_{s=0}=0.\ee
Assuming the foot point to
rotate at the Keplerian rate $\Omega=(GM/\po^3_0)^{1/2}$, we find
\be \p^2\Phi_e/\p s^2\vert_{s=0}=\Omega^2(\sin^2\theta-3\cos^2\theta), \ee
where $\theta$ is the angle between the field line and the $\po$ axis.
Thus the boundary occurs at a critical angle (Blandford and Payne, 1982):
\be \theta_c={\rm atan}(\sqrt 3)=60^\circ.\ee

\begin{figure}[tp]
\mbox{}\hfill\epsfysize6.2cm\epsfbox{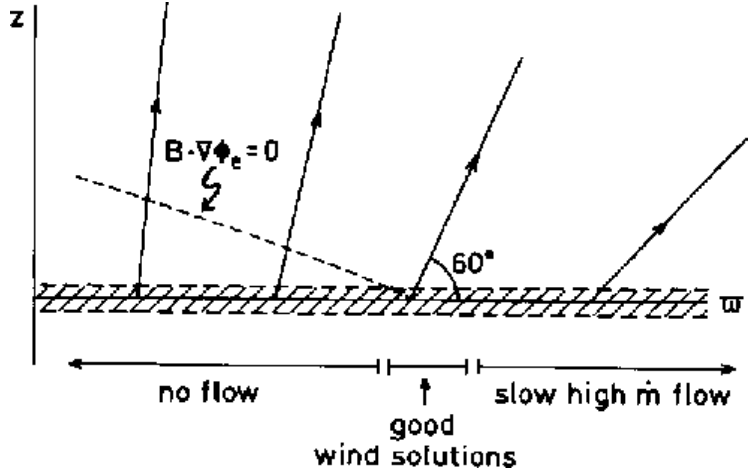}\hfill\mbox{}
%\picplace{6.2 cm}
\caption{\label{hsincl}
Launching conditions of a wind on field lines with varying inclination. The
maximum of the effective potential along field lines is shown by a broken
line. If the disk is cool, only a narrow range of inclinations around
$60^\circ$ yields `good' wind solutions. The slow, high $\dot m$ flows at low
inclination are problematic, see text.}
\end{figure}

The dependence on field line inclination is summarized in \fig{hsincl}.
On field lines more vertical than $60^\circ$, the situation is like in a
stellar wind: there is a potential barrier to overcome, and this requires the
existence of a hot atmosphere (temperatures comparable to the virial
temperature). For inclinations less than $60^\circ$, on the other hand, there
is no impediment to the flow at all, and one would expect a large mass flux.
This is, in fact, somewhat problematic, as discussed further in section
\ref{hscwd}, where the consequences of large mass fluxes are
investigated.
In between, there is a range in field inclinations (a narrow range
if the disk atmosphere is cool compared with the virial temperature), where a
reasonable mass flux results, and the magnetic wind theory works best. The
detailed solutions of K\"onigl (1989) are of this type. It has been argued
(Lubow et al. 1994) that the strong dependence of mass flux on inclination
makes the flows in this range unstable. If this is the case, stationary
magnetically accelerated flows may not exist, at least not from cool disks.
The conditions for stationary flows may be better in AGN. Here, the likely
presence of an ion supported torus (with ion temperature near the virial
temperature, Rees et al. 1982) near the black hole would allow a magnetically
generated flow from a wider range of field line inclinations.

From this discussion, it will be clear that the details of how the wind is
launched depend somewhat critically on things which are not presently known in
detail, namely the inclination of the field lines near the disk
surface and the presence or absence of a hot atmosphere.

\subsection{Geometry of the magnetic field near the disk}
\index{disks!magnetic fields}
In view of its importance for the wind launching problem, one would like to
know what determines the shape of the field lines near the disk surface. Since
the field is close to a potential field near the surface, its strength and
geometry in this region, including the inclination, is determined uniquely by
a boundary condition at the disk surface\footnote{A boundary condition at
large distance is also needed. Since the field strength in the disk is likely
to be much larger than in the interstellar medium, it is sufficient to take a
standard condition of vanishing field strength at infinity.}, namely the
vertical component of the field strength. There are two possibilities for the
origin of this field. If there is a dynamo process acting in the disk, one may
expect field strengths of the order of equipartition with the gas pressure, as
numerical simulations show (Hawley et al. 1995, Brandenburg et al. 1995).
These simulations also indicate that the field is created with small length
scales $L$, of the order of the disk thickness, in the radial direction, and
somewhat longer in the azimuthal direction. The potential field created by
such a small scale field decays with distance above the disk
like $\exp(-z/L)$. This would not be the ideal field for driving magnetic
winds.

Another possibility is that the field is not internally generated, but is due
to magnetic flux captured from the environment in during formation of the
disk, and advected and compressed by the accretion process. Poloidal flux
captured in this way cannot be destroyed by local processes in the disk, it
can only escape by diffusing radially outward. The field strength would then
be determined by the balance between outward diffusion and inward accretion
(van Ballegooijen 1989, Spruit 1994). In the absence of a theory for
(turbulent) diffusion in an accretion disk, it is not possible to predict with
any reliability what distribution of field strengths will result. Since all
energy densities in the disk increase inward however, including that of the
accretion flow, it is reasonable to assume that the balance will yield a field
with inward increasing strength. The field above the disk will then have a
shape like that suggested by \fig{hsincl}. In Spruit et al. (1995) we have
argued that fields of this configuration can be quite strong, with magnetic
energy densities exceeding the gas pressure, which would make them ideal for
the production of magnetic outflows.

\subsection{Poynting and kinetic energy fluxes}
\label{hssigma}
The wind carries both kinetic and magnetic energy. The asymptotic ratio of
these, at large distance, is a measure of how `magnetic' the wind is. The
Poynting flux, in the MHD approximation, is
\be {\bf S}={1\o 4\pi}({\bf v\times B}){\bf \times B}. \ee
The relevant component of $\bf S$ is that parallel to the poloidal flow. With
(\ref{hsvpr}) and (\ref{hsbpar}), this can be written as
\be S= {1\o 4\pi}\Omega({\bfg{$\po$}\times{\bf B})\times{\bf B}\cdot n}, \ee
where ${\bf n}$ is a unit vector along $\vp$. Working out the cross products:
\be S= \Omega\po {B_\rp B_\phi\o 4\pi}. \ee
Thus, the Poynting flux can be read as the work done by the rotation
against the magnetic torque. At large distance, the azimuthal velocity (in
the inertial frame) is small compared with $\Omega\po$, so that by
(\ref{hsbpar}) $B_p\approx B_\phi v_p/\Omega\po$, and $B_\phi\gg B_\rp$. Thus
\be
q\equiv{S\o K}\vert_\infty={B_\phi^2\o 2\pi\rho v_p^2}\vert_\infty=
2{v_{\rm A}^2\o v_\rp^2}\vert_\infty ,\label{hssig}
\ee
where $K$ is the kinetic energy flux ${1\o 2}\rho v^2$.
Many flows (an example is the cold Weber Davis model, \sect{hscwd}) have
their fast mode critical point at infinity so that $(v_{\rm
A}/v_\rp)_\infty=1$, and $q=2$. The magnetic and kinetic energy fluxes
are then comparable at infinity. Near the disk, the Poynting flux dominates.
Part of the Poynting flux is converted into a kinetic energy flux during the
acceleration process.

Relation (\ref{hssig}) is valid only if the field survives in a highly
wound-up form asymptotically. In section \ref{hskink} I will argue that
nonaxisymmetric instabilities are likely to
destroy at least part of the toroidal field. In reality, the magnetic
contribution to the energy flux may therefore be rather unimportant, and
$q\ll 1$ rather than of order unity. If this is the case, we have the
aesthetically pleasing result that the magnetic acceleration process, after
all its internal workings, produces a basically ballistic wind, which is only
moderately magnetic.

\subsubsection{electron-positron flows}
There is some observational evidence for outflows containing electron-positron
pairs (e$^\pm$) from relativistic objects. In the galactic center source
1E1740.7-2942 a positron annihilation feature has been observed (Churazov et
al. 1991, Churazov et al. 1994), and a transient
feature has been seen in X-ray Nova Muscae (Gil'fanov et al. 1991). Thus
there appear to be pair producing as well as jet producing black hole
candidates (though no case is known yet which combines both), and it is
natural to speculate that magnetic jets may exist that consist predominantly
of a pair plasma. The Blandford-Znajek mechanism is thought to produce such
flows (Blandford 1993 and references therein). Such a flow is technically not
different from the jets considered above, since the same MHD equations apply.
The main difference is that pairs may annihilate, thus removing mass and
inertia from the flow. This would tend to increase the relative importance of
the Poynting flux in the flow. At the same time, however, the Alfv\'en speed
would increase due to the decreasing mass density, and speed up the
instability of the toroidal field. As the mass disappears, the magnetic field
would therefore also disappear. In the nonrelativistic case, the flow would
then decay completely into photons. If the flow is relativistic, it is
possible that part of the energy of the decaying field is converted into a
flux of low-frequency electromagnetic waves. For recent speculations on this
topic, see Levinson and Blandford (1995). A magnetically driven pair plasma
flow from a pulsar has been invoked by Arons and collaborators (Gallant and
Arons 1994) for the Crab nebula.

\section{Cross-field balance}
\label{hscro}
The collimation of the flow is determined by the
force balance in meridional planes, in the direction perpendicular to the
field. The stationary, axisymmetric equation of motion that governs this
balance is called the Grad-Shafranov\index{Grad-Shafranov equation} or
Grad-Schl\"uter-Shafranov equation.
For our case of a magnetized flow it has a somewhat complicated form. Some
important aspects are discussed below, but for details I refer to
Heinemann and Olbert (1978) and Sakurai (1985). To begin with, note that the
solutions obtained in the above for a
fixed poloidal field are still valid for the full problem, provided we read
them as relations expressing the solution in terms of the (yet to be
determined) poloidal field.

The solutions of the azimuthal and longitudinal equations of motion are
(\ref{hssin}) [with (\ref{hspia})] and (\ref{hsbern}). Inserting these into
the
original equation of motion (\ref{hsmot}), we get the required expression for
the remaining, perpendicular, component. The result is (Heinemann and Olbert
1978, Okamoto 1975, 1992):
\bea
0&=&\nabla\psi\left\{{\rm div}\left[\left({\eta^2\o\rho}-{1\o 4\pi}\right)
{\nabla\psi\o\po^2}\right]- \rho\left(E^\pr-{1\o\gamma-1}{p\o\rho}
{K^\pr\o K}+\po^2\Omega\Omega^\pr\right) \nonumber\right. \\
& & \left.-{B^2\o\rho}\eta\eta^\pr-\po B_\phi\left[(\eta\Omega)^\pr
-{1\o\po^2}(\eta\Omega\ra^2)^\pr\right] \right\},\label{hsgs}
\eea
where a prime $^\pr$ denotes $\rd/\rd\psi$. It follows that the expression in
braces must vanish. This equation is to be read as a two-dimensional partial
differential equation for the stream function $\psi(\po,z)$ of the poloidal
field. Note that it is a somewhat implicit kind of equation, since it involves
the quantities $E(\psi),K(\psi),\ra(\psi)$ which are known in terms of $\psi$
only as solutions of eqs.\ (\ref{hssin}) and (\ref{hsbern}). Hidden in
(\ref{hsgs}) is the fact that it is singular at the Alfv\'en point. By working
out the coefficient of the highest (second) derivatives, one finds that it
vanishes at the Alfv\'en point, so a regularity condition must be applied
there. An additional complication
in solving the equation is that it is of mixed type, namely elliptic in some
parts of the $(\po,z)$ space and hyperbolic in others. Where the boundaries
are is found out by computing the characteristics of \eq{hsgs}. This
is conveniently done by inserting a short wave approximation:
\be \nabla\psi={\bf k}\psi, \ee
and keeping only the highest (quadratic) terms in $\bf k$. If $k_\parallel$
and $k_\perp$ are the components of $\bf k$ parallel and perpendicular to
$\bp$, this highest order treatment of the equation turns out to yield
\be
{k_\parallel^2\o k_\perp^2}={(v_\rp^2-v_{\rm cp}^2)( c_\rs^2+v_{\rm A}^2)\o
(v_p^2-v_{\rm sp}^2)(v_p^2-v_{\rm fp}^2)}, \label{hselhy}
\ee
where a new critical velocity $v_{\rm cp}$ has appeared:
\be v_{\rm cp}^2={ c_\rs^2v_{\rm Ap}^2\o c_\rs^2+v_{\rm A}^2}=
{v_{\rm sp}^2v_{\rm fp}^2\o v_{\rm sp}^2+v_{\rm fp}^2}. \label{hsvcp}\ee
As the flow accelerates it first meets an elliptic region
($k_\parallel^2/k_\perp^2<0$) for $v_\rp<v_{\rm cp}$, then a hyperbolic
region for $v_{\rm cp}<v_\rp<v_{\rm sp}$, another elliptic region $v_{\rm sp}<
v_\rp<v_{\rm fp}$, and finally another hyperbolic region for $v_\rp> v_{\rm
fp}$. The
significance of the critical velocity $v_{\rm cp}$ is seen by noting that it
is the speed of the {\em cusp} of an axisymmetric slow mode wave (Heinemann
and Olbert 1978). The surface on which $v_\rp=v_{\rm cp}$ is called the cusp
surface\index{cusp speed}. The cusp speed does not appear in the case of a
prescribed poloidal field, since the wave mode involved bends the poloidal
field lines.

The characteristics of \eq{hsgs} should not be confused with the
characteristics of a time dependent MHD problem. Though various wave speeds
appear, one is dealing with a time-independent flow. The ellipticity or
hyperbolicity of the problem refers to characteristics in the ($\po,z$) space,
not in an (${\bf r},t$) space. Physically, however, there is a clear relation
of the boundaries between elliptic and hyperbolic regions in the stationary
problem on the one hand, and wave speeds in a time dependent problem on the
other. This comes about because a wave in a frame comoving with
the fluid  appears as a stationary flow in the rest frame if the flow speed
just cancels the propagation of the wave. This happens at the critical
points.

The presence of 4 different regions poses practical problems when constructing
numerical solutions. The singular point at $v_\rp= v_{\rm Ap}$ has to be dealt
with, as well as the boundaries between elliptic and hyperbolic regions, since
different discretization schemes have to be used in each for numerical
stability. The lower boundary condition at the disk surface, together with the
regularity condition at the Alfv\'en surface act as the boundary conditions
for the elliptic regions (even though the boundaries of these regions do not
coincide with these surfaces!). The solutions in the inner and outer
hyperbolic regions are determined with values at the cusp and fast surfaces as
initial data, respectively. If $B_\phi\ll B_\rp$ near the sonic point, the
first hyperbolic region is quite narrow, and does not play an important role.
One can then regard the entire region inside fast magnetosonic surface as
elliptic, also in practical solution algorithms. This is the case when the
field is strong enough that `interesting' degrees of acceleration take place.
For discussions on numerical procedures, see Sakurai (1985, 1987) and
Camenzind (1987).

\section{The character of the wind at high and low $\dot m_{\rm w}$}
\label{hscwd}
A simple model which demonstrates important parts of the physics is that of
Weber and Davis (1967). I review it here in particular to discuss the
dependence of the wind problem on the mass flux. This also relates to the
question what happens to the flow when the inclination of the field line to
the disk surface is less than $60^\circ$ (cf. \sect{hslaunch}).

The model takes the poloidal field to be purely radial (in spherical
coordinates), and looks only at the equatorial plane (with respect to the
rotation axis)\footnote{A purely radial field smacks of monopoles. To remedy
this, the field below the equator is given the opposite sign of the field
above. The resulting configuration, with a current sheet at the equator, is
physically realizable. For the dynamics of the wind, the sign of the magnetic
field is unimportant. This is called the `split monopole' configuration.}.
Though the model was invented for stellar winds, it can also be applied to the
case of disk winds on low inclination field lines, nearly parallel to the disk
surface. To further simplify the problem, I ignore the thermal pressure
(`cold' limit), so that all
acceleration is magnetic.

Because he poloidal field is radial, its strength is given by
\be B_\rp=B_0(\po_0/\po)^2, \ee
where $\po_0$ is the foot point of the field line on the disk, which we assume
to rotate at the Keplerian rate $\Omega=(GM/\po^3)^{1/2}$.
For the analysis it is practical to normalize $\po$ and $\rho$ to their values
at the Alfv\'en point, by introducing the variables
\be x=\po/\ra; \qquad y=\rho/\rho_{\rm A}, \ee
and a normalized Bernoulli function
\be \tilde H=H{\ra\o GM}. \ee
Substituting (\ref{hssin}) into (\ref{hsbern}), the Bernoulli equation then
takes
the form
\be
\tilde H(x,y)={\beta\o 2 x^4y^2}+{\omega\o 2}{(x-1/x)^2\o(y-1)^2}
-{1\o x}-{\omega\o 2}x^2=E, \label{hshtil}
\ee
where
\be
\beta={B_0^2\po_0^4\o 4\pi GM\rho_{\rm A}\ra^3}=
\left[v_{\rm Ap}^2/{GM\o \po}\right]_{\rm A}, \qquad
\omega={\Omega^2\ra^3\o GM}=\left[\Omega^2\po^2/{GM\o\po}\right]_{\rm A}.
\ee
Since the sound speed vanishes, $v_\rp=0$ at the sonic point, and the gas
density diverges there. With (\ref{hssin}) the azimuthal velocity $v_\phi^\pr$
then vanishes also. Near the sonic point, the first two terms in
(\ref{hshtil})
describing the kinetic energy therefore vanish, and the only terms left are
the gravitational and centrifugal ones. The condition $\p_xH=0$ then yields
$\omega=x_s^{-3}$, in dimensional terms $GM=\Omega^2\po^3$. Thus, the sonic
point $x_s$ is at the foot point $x_0=\po_0/\po_{\rm A}$ of the field line,
and
\be x_0=\omega^{-1/3}. \label{hsx0}\ee
The value of the Bernoulli function is now also known,
\be E= -{1\o x_0}-{1\o 2}\omega x_0^2 = -{3\o 2}\omega^{1/3}. \ee
The relation between $\beta$ and $\omega$ follows from the fast point
condition. It turns out that the fast point is at infinity (Goldreich and
Julian 1970). We skip this part of the derivation.
One finds then that $x^2y$ remains finite at infinity, which
corresponds to the fact that the flow speed reaches a finite value at
infinity. Using the conditions $\p_yH=\p_xH=0$ for the fast point, and the
Bernoulli equation, all expanded for $x\rightarrow\infty$, the result is
\be (x^2y)_{\rm f}={2\o 3}-\omega^{-2/3}, \label{hsx2y}\ee
\be \beta=\omega({2\o 3}-\omega^{-2/3})^3. \label{hsbet}\ee
\begin{figure}[tp]
\mbox{}\hfill\epsfysize= 6cm \epsfbox{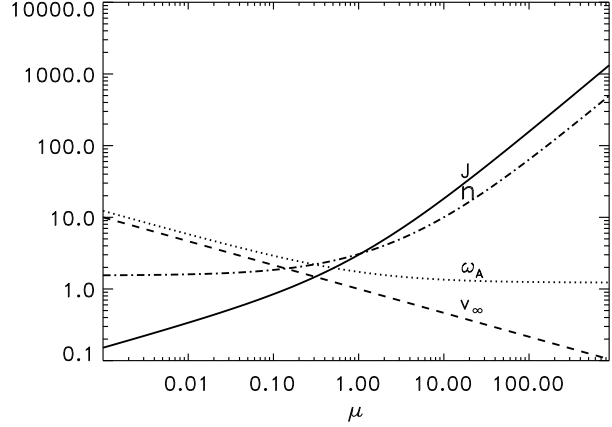}\hfill\mbox{}
\caption{\label{hsmup}
The cold Weber-Davis model, showing
Alfv\'en radius $\omega_{\rm A}=\po_{\rm A}/\po_0$, angular momentum flux
$J=\dot J/(\eta_*\Omega\po_0^2)$, asymptotic flow speed
$v_\infty/\Omega\po_0$, and field
angle $n=(B_\phi/B_{\rm r})_{\rm A}$ as functions of the mass flux.}
\end{figure}

We need to express this result in more physical terms. For a given field
strength and rotation rate, the external parameter determining the solution is
the density at the base of the flow or, equivalently, the mass flux. Consider
the mass flux as the external parameter. It is measured by the quantity
$\eta=\rho v_\rp/B_\rp$ [cf. (\ref{hseta})],
the mass flux `per field line'. This has the dimension of the square root of a
density. In fact, evaluating it at the Alfv\'en point,
\be \eta=\left({\rho_{\rm A}\o 4\pi}\right)^{1/2}. \ee
The quantity
\be \eta_*\equiv {B_0\o 4\pi\Omega \po_0} \ee
has the same dimension. It turns out to be the natural unit of mass flux in
the model. It increases with the field strength, reflecting the fact that a
stronger field is able to accelerate a larger mass flux to the same speed.
Defining a dimensionless mass flux $\mu$:
\be \mu=\eta/\eta_*, \label{hsmu}\ee
we want to express the results as functions of this dimensionless flux.
From the definition of $\beta$, and using (\ref{hsx0}), we find that
\be \beta=(\mu^2\omega)^{-1}. \ee
Hence with (\ref{hsbet}):
\be \omega=[{3\o 2}(1+\mu^{-2/3})]^{3/2}. \ee
The location of the Alfv\'en point is then
\be \ra/\po_0=\omega^{1/3}=[{3\o 2}(1+\mu^{-2/3})]^{1/2}. \ee
When the mass flux is small, $\mu\ll 1$, the Alfv\'en radius is far from the
origin of the flow. For large mass fluxes, $\mu\gg 1$, $\ra$ does not get
arbitrarily close to $\po_0$, but reaches the minimum value
\be \ra=\po_0(3/2)^{1/2}\quad (\mu\to\infty). \ee
Further quantities of interest are, for example, the angular momentum flux per
field line:
\be \dot J=\eta\Omega\ra^2=\eta_*\Omega\po_0^2\,\mu{3\o 2}(1+\mu^{-2/3}). \ee
This gives the angular momentum flux in terms of the mass flux and the
conditions at the base of the flow. The terminal speed of the flow follows
from (\ref{hseta}) and (\ref{hsx2y}):
\be {v_\infty\o \Omega\po_0}=(\beta\omega)^{1/6}=\mu^{-1/3}. \label{hsvinf}\ee
This demonstrates one of the most important properties of the magnetic
acceleration model: it can produce wind speeds that exceed the escape speed
$\Omega\po_0$ from the rotating object. In principle, it can accelerate a
sufficiently small mass flux to arbitrarily large speeds, though in practice
this ability is limited by the rather weak $1/3$ power in (\ref{hsvinf}).

For $\mu=1$, the final speed is just equal to the rotation velocity at the
base of the wind, and for large mass flux, the final speed becomes {\em
arbitrarily small}. What kind of flows are these massive but sluggish winds?
A good way to see this is by evaluating the pitch angle of the field at the
Alfv\'en point. The model gives for this
\be
\left({B_\phi\o B_\rp}\right)_{\rm A}=
(2-3\omega^{1/3}-\beta+\omega)^{1/2}\beta^{-1/2}.
\ee
The limiting forms are
\bea
\left({B_\phi\o B_\rp}\right)_{\rm A} &
     \approx & (19/8)^{1/2} \quad (\mu\ll 1) \\
                     &\approx   & 1.14\mu \quad (\mu\gg 1) \nonumber.
\eea
\begin{figure}[tp]
\mbox{}\hfill\epsfysize=5cm\epsfbox{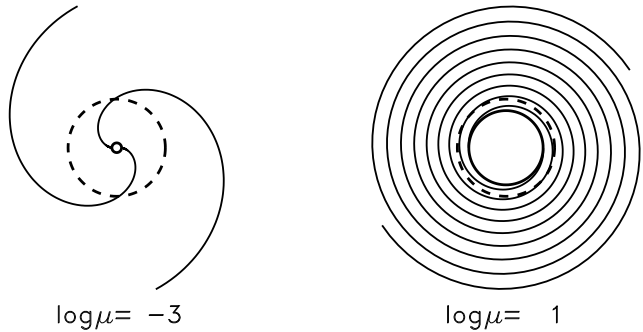}\hfill\mbox{}
\caption{\label{hsbfi} Shape of the field lines in a cold Weber-Davis model
for low (left) and high (right) mass loss cases. Dashed line: Alfv\'en radius.
}
\end{figure}
For small mass loss, the pitch angle of the field at the Alfv\'en radius
reaches a constant value which happens to be very nearly one radian. For large
$\mu$ however, the azimuthal field dominates over the poloidal field. This is
illustrated in \fig{hsbfi}, showing the shape of the field lines for a large
and a small value of $\mu$. The case $\mu\ll 1$ can be properly called a
centrifugally accelerated flow. Up to the Alfv\'en radius, the field lines are
not strongly bent, the flow corotates approximately, and the poloidal flow
speed can be found to good accuracy
from the effective potential. Beyond the Alfv\'en radius corotation fails, so
that the effective potential is not a good estimator for the flow speed any
more. For high mass loss, the situation is very different. Corotation now
fails right from the start, so that a strongly wound up $(B_\phi\gg B_p)$
field develops long before the Alfv\'en radius is reached. The flow is slowly
`pushed' outward by the pressure of the
toroidal field, with final speeds much less than $\Omega r_0$. Rather than
being `flung out', the flow is more a sequence of magnetostatic equilibria,
since the flow time scale $\po/v_\rp$ is much longer than the dynamical
time scales $(GM/\po^3)^{-1/2}$ and $\po/v_{\rm A}$. At the Alfv\'en point
($v_\rp=v_{\rm Ap}$), for example, the ratio of the flow and Alfv\'en time
scales is $v_{\rm A}/v_\rp= B_\phi/B_\rp\sim \mu$. This disparity of time
scales brings in the question of stability. If it is unstable, the highly
wound-up field will change on the short Alfv\'en time scale, interfering with
its pushing activity. For the low-$\mu$ centrifugal case, the field is not
strongly twisted, and stability is not an issue, until the flow reaches the
Alfv\'en radius. By then, most of the acceleration has already taken place. I
return to the question of stability in \sect{hskink}.

\subsection{Relativistic flows}
The relativistic case is somewhat outside the scope of this text. I will
discuss a few basic properties of the special-relativistic case, and refer to
Michel (1973), Goldreich and Julian (1970) and Li, Chiueh and Begelman (1992)
for details, and Bekenstein and Oron (1978), Okamoto (1978,1992) and Camenzind
(1987) for the general relativistic treatment. Also
left out is the Blandford-Znajek model for magnetic flows driven by the
rotation of a black hole. See Blandford (1993) and references therein.

In the nonrelativistic case, the only parameter determining the behavior of
the flow was the dimensionless mass loss rate $\mu$; the dependence on the
other physical parameters could be found by simple scalings. In the special
relativistic case, an additional parameter $w=\Omega\po_0/c$ appears because
the speed of light now fixes a velocity scale. Neglecting the gas pressure, a
Bernoulli equation can again be derived as in \sect{hstheor} (the derivation,
in the this case, is most easily done in the inertial frame). It can, as
before, be written in the form $H(\po,\rho)=E$, but analysing it is a bit more
complicated. In the extreme-relativistic limit, in which the asymptotic
Lorentz factor is large, the equivalent of relation (\ref{hsvinf}) becomes
\be \gamma_\infty={\Omega\po_0\o c}\mu^{-1/3}, \label{hsginf}\ee
(Michel, 1969), where $\mu$ is given by (\ref{hsmu}). As the mass flux is
decreased, and $\gamma_\infty$ increases, the Alfv\'en radius asymptotically
approaches the light cylinder radius $c/\Omega$. The high inertia of the flow
at large Lorentz factor ensures that $\po_{\rm A}$ always stays smaller than
$c/\Omega$.

Eq. (\ref{hsginf}) shows that the flow can in principle become relativistic
even when it is launched from a non-relativistic ($\Omega\po_0/c\ll 1$)
object, if the mass flux is low enough. In practice, however, the weak
dependence on $\mu$ means that Lorentz factors larger than a few can be
produced only by relativistic objects.

The cross field balance plays a more important role in the acceleration region
than in the non-relativistic case. Whereas in the non-relativistic case the
assumption of a fixed prescribed poloidal field is still fair near the
Alfv\'en surface, this is not the case for relativistic flows. In the extreme
relativistic limit, the inertial forces in the flow are so high near the
Alfv\'en surface that they bend the poloidal field lines into a nearly
horizontal shape at $\ra$ (Camenzind 1987).

\section{Collimation by `hoop stress'}
The curvature force exerted by the toroidal field compresses the field
configuration towards the axis. This effect becomes important only when the
toroidal field is comparable to or larger than the poloidal field, otherwise
the configuration is determined by the internal equilibrium of the poloidal
field. For low mass loss flows (cf \sect{hscwd}), collimation by the hoop
stress in the toroidal field therefore starts roughly at the Alfv\'en radius.
The effect was first observed in calculations of the solar wind by Suess and
Nerney (1975).

Note that collimation here is meant in `optical' sense: a flow is collimated
if the flow lines are parallel. This says nothing about the {\em width} of the
flow. For astrophysical jets, however, a collimated jet in practice is also
narrow. This is because the central engine is very small compared with the
scale of observed jets. An AGN jet for example that expands by a factor 1000
from its expected origin near the central black hole is still less than a
parsec across.

\begin{figure}[tp]
\mbox{}\hfill\epsfysize8cm\epsfbox{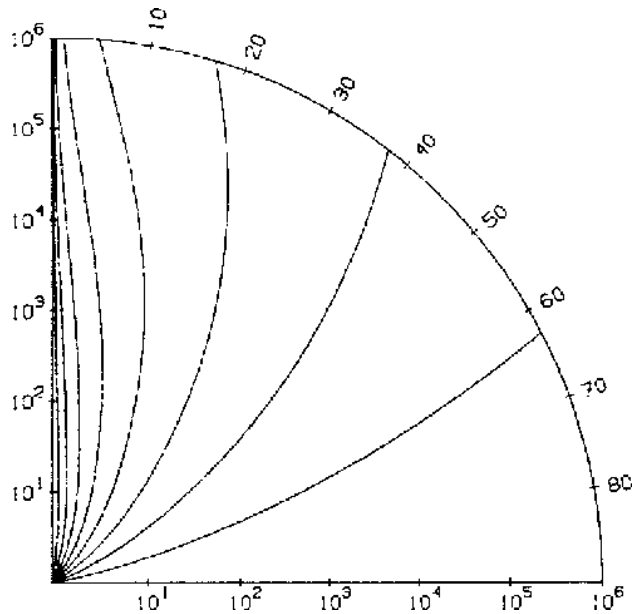}\hfill\mbox{}
%\picplace{8 cm}
\caption{\label{hssak}
Collimation of field lines in an initially spherical stellar wind. Radius
scale is logarithmic, in units of the Alfv\'en radius. All field lines
asymptotically become parallel to the axis. The logarithmic distance
scale distorts the field lines: on a linear scale the distance of each field
line from the axis increases monotonically.
From Sakurai (1985).}
\end{figure}

The asymptotic collimation of an initially radial flow is illustrated in
\fig{hssak}. In this model (Sakurai, 1985) a flow is launched
spherically symmetric on a `split monopole' field. After passing through
the Alfv\'en point, the field becomes predominantly toroidal, and this causes
the flow to become perfectly collimated (all flow lines parallel to the axis)
no matter how small the rotation rate of the star. The rate at which this
collimation takes place, however, is slow, requiring several orders of
magnitude in distance. For ordinary stellar rotation rates, the distance
needed for full collimation is unrealistically large: the flow reaches its
interstellar termination shock well before being collimated.

In addition to the perfectly collimated flows like Sakurai's, solutions have
been found that are asymptotically uncollimated, in spite of the hoop stress.
If the field configuration is such that the field lines diverge sufficiently
rapidly near the Alfv\'en surface (faster than a purely radial field), the
fast mode point is located at only a few Alfv\'en radii, and the flow remains
uncollimated (Begelman and Li 1994). Depending on the conditions in the
accelerating region, it seems one either gets an asymptotically fully
collimated flow (the `cylindrical' case) or an uncollimated, space-filling
flow (a `conical' flow, see also Sauty and Tsinganos 1994, Nitta 1994,
Tomimatsu 1994, Heyvaerts and Norman 1996). In the latter case, the asymptotic
ratio $q$ of magnetic to kinetic energy flux is smaller than in the
cylindrical case [where $q=2$, cf.\ \sect{hssigma}].

The process of collimation by hoop stress is a natural consequence in
axisymmetric rotating winds, and can be computed in detail. The completeness
and accuracy suggested by such computations, however, is somewhat misleading
because they depend very heavily on the assumption of axisymmetry. If regions
of predominantly toroidal field are as unstable as toroidal fields elsewhere
in the universe and the laboratory, a significant revision of our picture of
the collimation of magnetic winds is in order (\sect{hssigma}).

\section{Kink instability}
\label{hskink}
\begin{figure}[tp]
\mbox{}\hfill\epsfysize5.7cm\epsfbox{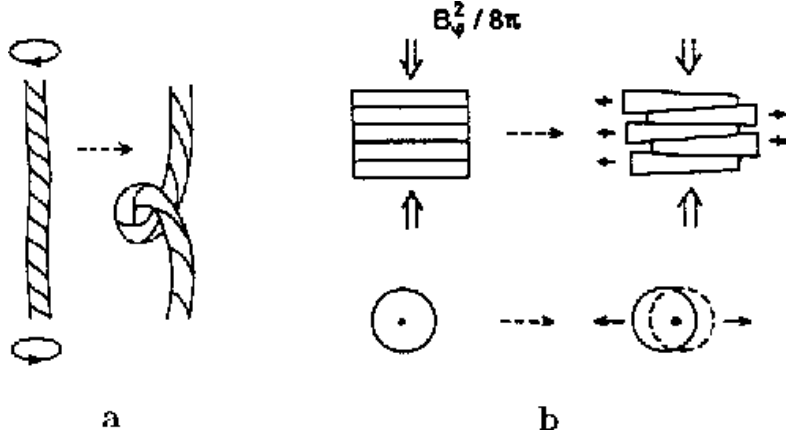}\hfill\mbox{}
%\picplace{5.7 cm}
\caption{\label{hskin}
Sketch of the kink instability mechanism. a: in analogy with an
overtwisted rubber band. b: when the azimuthal field dominates, the
instability is like that of a stack of deformable disks under compression.}
\end{figure}

\index{instability!kink}
In the previous section we found that a predominantly toroidal field develops
outside the Alfv\'en surface. In high mass loss flows, it develops also inside
the Alfv\'en radius. Consider first the case of a low-$\mu$ flow, outside the
Alfv\'en surface. Assume that the flow is well collimated, and move
into a frame comoving with the flow. In this frame, we see a toroidal field,
slowly decreasing in time by the expansion of the flow. A predominantly
toroidal field, however, is violently unstable to kink instabilities: such a
configuration is equivalent to the linear pinch (e.g.\ Roberts 1967, Parker
1979, Bateman 1980). The mechanism of the instability is illustrated in
\fig{hskin}. An initially axial, untwisted, magnetic field  is wound up
and becomes unstable when the azimuthal becomes larger than the axial field
strength. This is akin to the instability of a twisted rubber band
(\fig{hskin}a). Instability sets in when the axial tension vanishes. Denoting
by
$B_z$ and $B_\phi$ the axial and azimuthal components of the field, the axial
component of the stress is $(-B_z^2+B_\phi^2)/8\pi$. The first term is the net
magnetic tension due to the axial field, and is stabilizing; it likes to keep
field lines straight. The second term, equal to the magnetic pressure exerted
by the azimuthal component, is positive, expansive. When the pressure becomes
larger than the tension, some of the energy put in by the twisting is released
by a kink. Each kink reduces the number of windings by one, at the expense of
increasing the energy in the axial field by lengthening axial field lines
somewhat\footnote{The condition $B_\phi>B_z$ can underestimate the
degree of instability. A cylindrical field configuration typically becomes
unstable already when it is twisted by more than one full turn, independent of
the distance between the surfaces at which the twist is applied
(Kruskal-Shafranov condition). In our case, this is not relevant, however,
because $B_\phi/B_\rp$ increases with distance in such a way that the number
of turns in the field is always less than one at the point where $B_\phi$
first exceeds $B_\rp$.}.

The kink instability is a transition to a nearby equilibrium of lower energy,
i.e. the instability saturates at a finite amplitude. This is because the
amount of azimuthal field energy that can be released is finite, while the
energy expended lengthening the the axial field increases indefinitely with
the amplitude of the perturbation. In a predominantly azimuthal field, the
instability can also be visualized as shown in \fig{hskin}b. A stack of
deformable disks (think of the disks in your spinal column, for example) is
compressed (by the pressure of the azimuthal field). By slipping sideways and
deforming somewhat, the disks can release some of the pressure, at the expense
of the integrity of the stack. In stellar interiors, such configurations are
also known to be highly unstable in spite of the presence of a stabilizing
thermal buoyancy force (Tayler 1980, Pitts and Tayler 1985).

In these descriptions of the instability, the field is treated as if embedded
in a neutral medium, and the instability is a so-called `external' kink. In
practice, the flow could be surrounded by, or itself surround, a less twisted
field configuration. This has a stabilizing effect. Conditions for
instability in this case (the `internal' kink) are somewhat more complicated.
For more on the subject see, e.g.\ Bateman (1980, Ch. 6)\footnote{In
Bateman, and in the controlled fusion literature, the use of the words
`poloidal' and `toroidal' in case of cylindrical fields is opposite to our
usage; this has to do with the torus geometry assumed there.}.

Kink instability develops on a time scale $\po/v_{\rm A\phi}$, the Alfv\'en
travel time across the flow, based on the azimuthal field strength. The moment
that instability takes place, the azimuthal field providing the
collimating\index{outflows!collimation}\index{jets!collimation}
hoop stress is reduced (see also Eichler, 1993). The energy involved goes into
a less ordered field component which, if anything, adds {\em outward} magnetic
pressure instead of an organized force towards the axis. Thus, the collimating
effect of $B_\phi$ decays at the same rate as the instability takes place. The
reduction of the collimating hoop stress has the strongest effect in the most
collimated flows; this is seen as follows. If the collimation angle is
$\theta$, the radial expansion speed of the jet is $v_\po=\theta v_\rp
\approx \theta v_{\rm A}$, which is small compared with the Alfv\'en speed.
Hence the
instability has ample time to act as the jet moves outward. The effect of the
instability would be less dramatic close to the Alfv\'en radius. Choudhuri and
K\"onigl (1986) have proposed that kink instability near the Alfv\'en radius
may be responsible for some of the alignment anomalies seen in jets at the
VLBI\index{jets!VLBI} scale.

It takes longer than the instability time scale to dissipate the
disorganized field component it produces (this is related to the known slow
dissipation of magnetic helicity\index{helicity!magnetic}, and is seen also in
numerical simulations, e.g.\ Galsgaard 1995). This dissipation, however,
eventually leads to a
reduction of the field strength compared with the standard axisymmetric jet. A
second consequence of kink instability is therefore that the ratio of magnetic
to kinetic energy flux in the jet becomes less than unity (see
\sect{hssigma}). Since the Alfv\'en speed is lower, the fast mode critical
point is closer to the source, perhaps at only a few Alfv\'en radii. Most of
the observed jet would then be outside the fast mode point, and kinetic energy
dominated. In short: the jet behaves like a ballistic flow, like a water jet
from a fire hose. This would simplify the magnetic jet picture considerably:
though the acceleration process is intensely magnetic, it would eventually
produce a ballistically moving jet in which magnetic stresses are a secondary
factor as far as the dynamics is concerned.

\begin{figure}[tp]
\mbox{}\hfill\epsfysize5cm\epsfbox{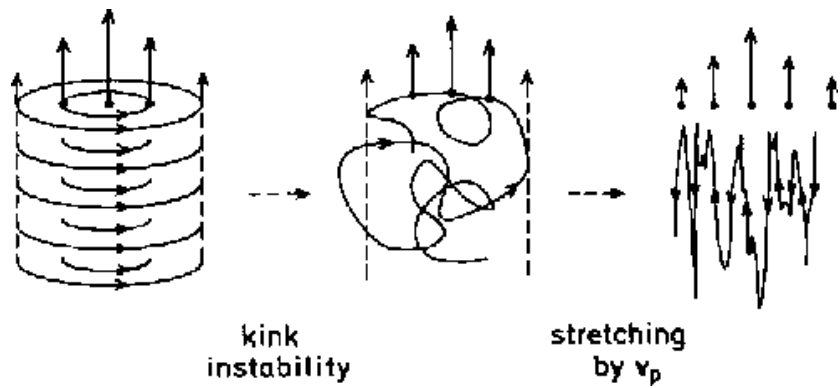}\hfill\mbox{}
%\picplace{5 cm}
\caption{\label{hsstretch}
Production of a longitudinal field by kink instability. The longitudinal flow
is taken to depend on distance from the jet axis. Initially parallel to
magnetic surfaces, instability forces the flow to cross the displaced
field lines. The differential flow speed stretches the displaced azimuthal
field lines along the axis.}
\end{figure}

Some observational evidence for the action of kink instabilities may be the
fact that the magnetic field tends to be parallel to jet axis, at least in the
faster (type II) jets (Bridle and Perley, 1984). If the flow speed along the
jet is not exactly  uniform over its cross section, the irregularities in the
field produced by the instability will be stretched along the jet axis, see
\fig{hsstretch}. The strength of this longitudinal field will be comparable to
the kinetic energy of {\em differential} velocity across the jet. This field
will have many small scale reversals of direction, explaining why the total
poloidal magnetic flux inferred from observations (which are not sensitive to
the direction of the field lines) is much larger than can be easily
accomodated in the accelerating region. These observational indications can
equally be explained by stretching of the field by interaction with an
external medium, but irregularities produced internally by kinking have the
advantage that they will also work in the absence of any interaction with the
surroundings.

\section{Poloidal collimation}
\begin{figure}[tp]
\mbox{}\hfill\epsfysize=5cm\epsfbox{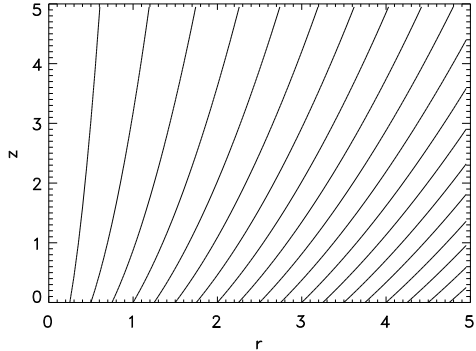}\hfill\mbox{}
\caption{\label{hsparf}
Field lines of a potential field produced by a magnetic field strength varying
as $(r^2+1)^{-1/2}$ on the disk surface. The collimating shape of the field
lines is due to the magnetic flux in the outer parts of the disk.}
\end{figure}

\index{outflows!collimation}\index{jets!collimation}
In addition to `toroidal' collimation by hoop stresses, a poloidal magnetic
field surrounding the jet can also be a collimator. This is likely to be a
powerful effect (Blandford 1993, Spruit 1994) if the disk is `Large' (where by
`Large' I mean extending over a significant number of decades in radius). As
an example to demonstrate this, assume that the vertical field strength at the
surface of the disk is of the form
\be B_z\sim (r^2/r^2_i+1)^{-\nu/2}, \label{hsnu}\ee
where $r_i$ is the inner edge of the disk and we take $\nu$ to be between 0
and 2. Then the field strength is largest in the inner parts of the disk, but
the magnetic flux $\int rB_z\rd r$ is dominated by the outer regions of
the disk. This is a reasonable situation to expect for the field in a disk. A
radially selfsimilar disk, for example (Blandford and Payne 1982), has
$\nu=5/4$. In the inner accelerating region of the flow such a field is close
to the potential field given by the distribution of flux on the disk surface.
It turns out that for a distribution like (\ref{hsnu}), the field lines have a
nice, naturally collimating, shape. An example is shown in \fig{hsparf}
(Spruit 1994), which shows the field lines for the case $\nu=1$. In this
case, the field lines are parabolas, hence their collimation becomes {\em
perfect} at large distance.
Of course, the field stops to have this shape near the Alfv\'en surface and
beyond, and at distances where the finite size of the disk becomes noticeable.
We can derive a maximum degree of collimation from these ideas, as follows
(Spruit, Foglizzo and Stehle 1996). The best collimation is obtained when the
Alfv\'en surface is at a distance of the order of the disk size, but not
further. At larger distances than this, the field more resembles that of a
dipole, and does not have any collimating properties. For a field of the form
(\ref{hsnu}) one finds (for the case $\nu=1$) that the angle of the field
lines with the axis, at the Alfv\'en surface, is
\be \theta_{\rm min}\approx (r_{\rm i}/r_{\rm d})^{1/2}, \ee
where $r_{\rm i}$ and $r_{\rm d}$ are the inner and outer radii of the disk.
If we assume that no further collimation takes place beyond the Alfv\'en
surface, for example because of the kink instabilities discussed above, this
angle is also the minimum opening angle of the jet. If the Alfv\'en distance
is significantly smaller or larger than $r_{\rm d}$, the collimation is worse.

We can compare these minimum angles of collimation with conditions expected
for various kinds of disk. This is shown in \tab{hscol}.
\begin{table}[bp]
\caption{\label{hscol}
Typical disk dimensions and minimum collimation angle for poloidal
collimation, for different kinds of accretion disk systems.}
\begin{tabular}{lllr}
  & $r_{\rm i}$ & $r_{\rm d}$ & $\theta_{\rm min}$ \\
protostars & 0.01 AU & 100 AU & 0.01 \\
LMXB & 10 km & $10^5$km & 0.01 \\
AGN & 1 AU & $>10^4$AU & $<0.01$ \\
CV & $10^4$km & $2\,10^5$km & 0.2 \\
R Aqr & $10^4$km & $>10^8$ km(?) & $<0.01$
\end{tabular}
\end{table}
This shows that poloidal collimation is capable of explaining opening angles
of less than a degree in most systems, with the notable exception of
Cataclysmic Variables. And, in fact, no CV is known to produce a jet, though
there is evidence for outflows from many such objects (section
\ref{hsobs}). I interpret this as a good case for the importance of poloidal
collimation. A nice test case in this context is R Aqr. It consists of a white
dwarf accreting from a giant companion, demonstrating that it is not the white
dwarf nature of the primaries in CV that prevents them from having jets.
Because of its very long orbital period, the disk in this system is probably
several orders of magnitude larger than the disks in CV.

Confirmation that the relative disk size $r_{\rm d}/r_{\rm i}$ is important
may perhaps be found in binary protostars. Several cases are now known (see
Matthieu, 1996) of relatively close binary protostars where at
least one of the stars has a disk. The maximum size of such a disk can not be
much larger than the tidal radius, something of the order of 1/3 of the
orbital separation. The prediction is then that such disks do not produce
well-collimated jets if $r_\rd\lapprox 30R_*$, i.e. if the orbital separation
is less than about 0.5AU.

\leftline{\bf acknowledgements}
This work was done in the Human Capitical and Mobility
network `Accretion in Close binaries' (CHRX-CT93-0329). I thank Rudi Stehle
and Thierry Foglizzo for their comments on an earlier version of this text.

\vspace{1 cm}
\leftline{\bf References}
\noindent \hangindent=2.5em
Bateman, G.: 1980,  {\em MHD Instabilities}, Cambridge
(Mass.): MIT press

\noindent \hangindent=2.5em
Begelman, M.C., Blandford, R.D. and Rees, M.J.: 1984, 
{\em  Rev. Mod. Phys.} {\bf 56}, 255

\noindent \hangindent=2.5em
Begelman M.C. and Li Z.-Y., 1994, \apj{426}, 269

\noindent \hangindent=2.5em
Bekenstein, J.D. and Oron, E.: 1978, 
{\em Phys. Rev D} {\bf 18}, 1809

\noindent \hangindent=2.5em
Bisnovatyi-Kogan, G. and Ruzmaikin, A.A.: 1976, 
{\em Astrophys Sp. Sci.} {\bf 42}, 401

\noindent \hangindent=2.5em
Blandford, R.D.: 1976, 
\mnras{176}, 465

\noindent \hangindent=2.5em
Blandford, R.D. and Payne, D.G.: 1982,  
\mnras{199}, 883

\noindent \hangindent=2.5em
Blandford, R.D.: 1993, 
in D. Burgarella, M. Livio and C. O'Dea, eds., {\em Astrophysical Jets},
Cambridge: Cambridge University Press, 15.

\noindent \hangindent=2.5em
Brandenburg, A., Nordlund, {\AA}., Stein, R.F. and Torkelsson, U.: 1995,
 \apj{446}, 741

\noindent \hangindent=2.5em
Bridle, A.H. and Perley, R.A.: 1984, \annrev{22}, 319

\noindent \hangindent=2.5em
Burgarella, D. and Paresce, F.: 1992  \apj{389}, L29, errratum in
{\bf 395}, 123

\noindent \hangindent=2.5em
Camenzind, M.: 1987, 
\aaa{184}, 341

\noindent \hangindent=2.5em
Cao, X.-W. and Spruit H.C.: 1994, 
\aaa{287}, 80

\noindent \hangindent=2.5em
Choudhuri, A.R. and K\"onigl, A.: 1986  \apj{310}, 96

\noindent \hangindent=2.5em
Churazov, E., Sunyaev, R.A. and Gil'fanov, M., et al.: 1994,
\apjs{92}, 381

\noindent \hangindent=2.5em
Churazov, E., Gil'fanov, M., Sunyaev. R.A., et al.: 1993,
\apj{407}, 752

\noindent \hangindent=2.5em
Cisowski, S.M. and Hood, L.L.: 1991, in C. P.Sonnett, M.S. Giampapa, M.S.
Matthews, eds., {\em The Sun in Time}, Tucson: Univ. of Arizona
Press, p.761

\noindent \hangindent=2.5em
Dougherty, S.M., Bode, M.F., LLoyd, H.M., Davis, R.J. and Eyres, S.P.: 1995

\mnras{272}, 843

%\noindent \hangindent=2.5em
%Draine, B.T.: 1983,  \apj{270}, 519

\noindent \hangindent=2.5em
Drew, J.E. and Verbunt, F.: 1988, 
\mnras{234}, 341

\noindent \hangindent=2.5em
Eichler, D.: 1993,  \apj{419}, 111

\noindent \hangindent=2.5em
Foukal, P.V.: 1990, 
{\em Solar Astrophysics}, Wiley, New York, Chapter 12.3

\noindent \hangindent=2.5em
Gallant, Y.A. and Arons J.: 1994, \apj{435}, 230

\noindent \hangindent=2.5em
Galsgaard, K.: 1995, {\em J. Gephys. Res.} in press

\noindent \hangindent=2.5em
Gil'fanov, M., Sunyaev, R.A., Churazov, E. et al.: 1991, {\em Pis'ma
Astron. Zh.} {\bf 17}, 1059, translation in {\em Sov. Astron. Lett.} {\bf 17},
437 (1992)

\noindent \hangindent=2.5em
Goldreich, P. and Julian, W.H.: 1969 
\apj{157}, 869

\noindent \hangindent=2.5em
Goldreich, P. and Julian, W.H.: 1970 
\apj{160}, 971

\noindent \hangindent=2.5em
Hack, M. and La Dous, C., 1993, 
{\em Cataclysmic Variables and related objects}, NASA SP-507, Washington D.C.:
US Printing Office

\noindent \hangindent=2.5em
Hawley, J.F., Gammie, C.F. and Balbus, S.A.: 1995, 
\apj{440}, 742

\noindent \hangindent=2.5em
Heinemann, M. and Olbert, S.: 1978, 
{\em J. Geophys. Res.} {\bf 83}, 2457

\noindent \hangindent=2.5em
Heyvaerts, J. and Norman, C.A.: 1989, 
\apj{347}, 1055

\noindent \hangindent=2.5em
Heyvaerts, J. and Norman, C.A.: 1996, 
preprint

\noindent \hangindent=2.5em
Hjellming, R.M. and Rupen, M.P.: 1995  \nature{375}, 464

\noindent \hangindent=2.5em
Icke, V., Balick, B. and Frank, A.: 1992, 
\aaa{253}, 224

\noindent \hangindent=2.5em
Jeffrey, A. and Taniuti, T.: 1964, 
{\em Non-linear Wave propagation}, New York: Academic Press.

\noindent \hangindent=2.5em
K\"onigl, A.: 1989, 
\apj{342}, 208

\noindent \hangindent=2.5em
K\"onigl, A. and Ruden, S.P.: 1993, 
in {\em Protostars and Planets III}, eds. E.H. Levy,\& J.I. Lunine, Tucson:
Univerity of Arizona Press, 641

\noindent \hangindent=2.5em
Levinson, A. and Blandford, R.D.: 1995, preprint 9506137, Los Alamos/Sissa
e-print service

\noindent \hangindent=2.5em
Lewin, W.H.G., van Paradijs, J. and van den Heuvel, E.P.J., eds.: 1995,
 {\em X-ray Binaries}, Cambridge Univ. Press, Cambridge

\noindent \hangindent=2.5em
Li, Z., Chiueh, T. and Begelman, M.C.: 1992,  \apj{394}, 459

\noindent \hangindent=2.5em
Lichn\'erowicz, A.: 1967, 
{\em Relativistic hydrodynamics and magnetohydrodynamics}, New York: Benjamin.

\noindent \hangindent=2.5em
Lovelace, R.V.E.: 1976  \nature{262}, 649

\noindent \hangindent=2.5em
Lovelace, R.V.E., Wang, J.C.L. and Sulkanen, M.E.: 1987,  \apj{315},
504

\noindent \hangindent=2.5em
Lubow, S., Papaloizou, J.C.B. and Pringle, J.E: 1994, 
\mnras{268}, 1010

\noindent \hangindent=2.5em
Lubow, S. and Spruit, H.C: 1995, 
\apj{445}, 337

\noindent \hangindent=2.5em
Mathieu, R.: 1996 in  R.A.M.J. Wijers,
C.A. Tout and J.E. Pringle (eds.) {\it Physical processes in Binaries}, Kluwer
Dordrecht, (NATO ASI series).

\noindent \hangindent=2.5em
Mestel, L.: 1968  \mnras{138}, 359

\noindent \hangindent=2.5em
Michel, F.C.: 1969 
\apj{158}, 727

\noindent \hangindent=2.5em
Michel, F.C.: 1973 
\apj{180}, L133

\noindent \hangindent=2.5em
Mirabel, I.F., Cordier, B., Paul and J., Lebrun, F.: 1992,
\nature{358}, 215

\noindent \hangindent=2.5em
Mirabel, I.F. and Rodriguez, L.F.: 1994  \nature{371}, 46

\noindent \hangindent=2.5em
Nitta, S.-y.: 1994, \pasj{46}, 217

\noindent \hangindent=2.5em
Okamoto, I.: 1975, 
\mnras{173}, 357

\noindent \hangindent=2.5em
Okamoto, I.: 1978: 
\mnras{185}, 69

\noindent \hangindent=2.5em
Okamoto, I.: 1992, 
\mnras{254}, 192

\noindent \hangindent=2.5em
Parker, E.N.: 1963,  {\em Interplanetary Dynamical Processes}, New
York: Wiley.

\noindent \hangindent=2.5em
Parker, E.N.: 1979, 
{\em Cosmical Magnetic Fields: Their Origin and Activity}, Oxford: Clarendon,
Ch. 9

\noindent \hangindent=2.5em
Pitts, E. and Tayler R.J.: 1985,  \mnras{216}, 139

\noindent \hangindent=2.5em
Plumpton, C. and Ferraro, V.C.A: 1966, 
{\em Introduction to Magneto-fluid dynamics}, Oxford: Clarendon.

\noindent \hangindent=2.5em
Pudritz R.E. and Norman C.A.: 1986  \apj{301}, 571

\noindent \hangindent=2.5em
Rees, M.J., Begelman, M.C., Blandford, R.D. and Phinney, E.S.: 1982, 
\nature{295}, 17

\noindent \hangindent=2.5em
Roberts, P.H.: 1967 {\em An Introduction to Magnetohydrodynamics}, London:
Longmans, p.235

\noindent \hangindent=2.5em
Sakurai, T.: 1985, 
\aaa{152}, 121

\noindent \hangindent=2.5em
Sakurai, T.: 1987, 
\pasj{39}, 821.

\noindent \hangindent=2.5em
Sauty, C. and Tsinganos, K.: 1994, 
\aaa{287}, 893

\noindent \hangindent=2.5em
Schatzman, E.: 1962  {\em Ann. Astrophys.} {\bf 25}, 18

\noindent \hangindent=2.5em
Spruit, H.C.: 1994, in 
{\em Cosmical Magnetism}, ed.\ D.\ Lynden-Bell, Dordrecht: Kluwer, p. 33

\noindent \hangindent=2.5em
Spruit, H.C., Stehle R.and Papaloizou, J.C.B.: 1995, 
\mnras{275}, 1223

\noindent \hangindent=2.5em
Spruit, H.C., Foglizzo, T. and Stehle R.: 1996, 
\aaa{} in preparation.

\noindent \hangindent=2.5em
Stewart, R.T., Caswell, J.L., Haynes, R.F., Nelson, G.J.: 1993, 
\mnras{261}, 593

\noindent \hangindent=2.5em
Strom, R.G., van Paradijs, J., van der Klis, M.; 1989  \nature{337},
234

\noindent \hangindent=2.5em
Suess, S.T., Nerney, S.F.: 1975, 
\sp{40}, 487

\noindent \hangindent=2.5em
Sunyaev, R.A., Kaniovskii, A.S., Efremov, V.V. et al.: 1991,  {\em
Pis'ma Astron. Zh.} {\bf 17}, 291, translation in {\em Sov. Astron. Lett.}
{\bf 17}(2), 123 (1991)

\noindent \hangindent=2.5em
Tayler, R.J.: 1980  \mnras{191}, 151

\noindent \hangindent=2.5em
Thorstensen, J.R., Ringwald, F.A., Wade, R.A., Schmidt, G.D., Norsworthy,
J.E.: 1991, 
\aj{102}, 272.

\noindent \hangindent=2.5em
Tomimatsu, A.: 1994, \pasj{46}, 123

\noindent \hangindent=2.5em
van Ballegooijen, A.A.: 1989, in {\em Accretion Disks and Magnetic Fields in
Astrophysics}, ed. G. Belvedere, Dordrecht: Kluwer, p.99

\noindent \hangindent=2.5em
Weber, E.J., Davis, L.: 1967  \apj{148}, 217

%Woods J.A., Drew J.E., Verbunt F. 

\end{document}